\documentclass[aps,prd,twocolumn,superscriptaddress,nofootinbib]{revtex4-2}
\usepackage{amsmath,amssymb,graphicx,bm}
\usepackage{tikz}
\usepackage{pgfplots}
\pgfplotsset{compat=1.9}
\usetikzlibrary{patterns,arrows.meta}
\definecolor{figblue}{RGB}{31,78,121}
\makeatletter
\ifcsname prep@math@patch\endcsname
\renewenvironment{widetext}{%
  \par\ignorespaces
  \onecolumngrid
  \vskip10\p@
  \prep@math@patch
}{%
  \par
  \vskip8.5\p@
  \twocolumngrid\global\@ignoretrue
  \@endpetrue
}%
\fi
\makeatother
\definecolor{figgreen}{RGB}{46,139,87}
\definecolor{figred}{RGB}{176,48,48}
\usepackage[colorlinks=true,linkcolor=blue,citecolor=blue,urlcolor=blue]{hyperref}

\newcommand{\eps}[1]{\varepsilon_{#1}}
\newcommand{\stw}{\sin^2\theta_{12}}
\newcommand{\sth}{\sin^2\theta_{13}}
\newcommand{\stt}{\sin^2\theta_{23}}

\begin{document}

\title{The $\pi/12$ model: trimaximal first-column lepton mixing with charged-lepton $\mu$--$\tau$ breaking}

\author{Vernon Barger}
\affiliation{Department of Physics, University of Wisconsin--Madison, Madison, WI 53706, USA}

\date{2026-09-08}

\begin{abstract}
The first column of the lepton mixing matrix carries two independent magnitude conditions: the trimaximal norm $|U_{e1}|^2 = 2/3$ and the $\mu$--$\tau$ balance $|U_{\mu1}| = |U_{\tau1}|$. We parameterize their violation by $\eps{1} \equiv \tfrac{3}{2}|U_{e1}|^2 - 1$ and $\eps{2} \equiv 3(|U_{\tau1}|^2 - |U_{\mu1}|^2)$ and show that current data treat the two differently in central value: with the first JUNO measurement included, $\eps{1} = +0.014 \pm 0.010$ is consistent with zero, while $\eps{2} = +0.29^{+0.07}_{-0.13}$ is large but disfavors zero at only $1.5\sigma$ on the profile likelihood of the global fit, the phase that drives it being still poorly measured. This asymmetry selects a specific framework, which we call the $\pi/12$ model. Its neutrino sector is $\text{TM}_1$, supplemented by the electron-row condition $|U_{e2}|/|U_{e3}| = 2+\sqrt{3}$ that fixes $\sth = (2-\sqrt{3})/12 = 0.02233$ and $\stw = 0.31811$. A charged-lepton 2--3 rotation $R_{23}(\theta_e,\alpha)$ then leaves the electron row exactly invariant and generates $\eps{2} = \sin 2\theta_e \cos\alpha$. In a concrete $S_4 \times C_4 \times C_3 \times C_2$ realization, the rotation arises from a single additional flavon whose alignment forces $\alpha = 0$; a natural rotation angle $\theta_e \simeq 2^\circ$ then reproduces the observed departure of $\theta_{23}$ from maximal while pinning the CP phase to $\delta = 272^\circ \pm 2^\circ$, near-maximal CP violation with $\eps{2} = -0.066$. The framework thereby converts the current $\theta_{23}$--$\delta$ measurements into a sharp test: the predicted $\delta$ sits $1.7\sigma$ from the present central value on the same profile likelihood, the predicted $\eps{2}$ has the opposite sign to the measured one, and the Hyper-Kamiokande design precision of $20^\circ$ on $\delta$ resolves the question at the $3\sigma$ level. The neutrino-sector mass predictions are inherited unchanged, $\Sigma m_\nu = 65.6\ \text{meV}$ and $m_{\beta\beta} = 5.4\ \text{meV}$, placing the model at the current DESI bound in $\Lambda$CDM.
\end{abstract}

\maketitle

\section{Introduction}
\label{sec:intro}

The first JUNO measurement, $\stw = 0.3092 \pm 0.0087$ from 59.1 days of reactor data \cite{JUNO:2025}, has changed the status of constant mixing patterns from a topic of model preference to one of experimental discrimination. Against the global fit that incorporates it \cite{NuFIT61}, the tribimaximal value $\stw = 1/3$ \cite{HPS:2002} is disfavored at $3.7\sigma$ and the $\text{TM}_2$ prediction $\stw = 0.3410$ at $4.8\sigma$, while the $\text{TM}_1$ prediction $\stw = 0.3181$ remains compatible at $1.4\sigma$ \cite{Zhang:2025}. Among the one-parameter deformations of tribimaximal mixing, the pattern that preserves the first column,
\begin{equation}
|U_{e1}| = \tfrac{2}{\sqrt6}, \qquad |U_{\mu1}| = |U_{\tau1}| = \tfrac{1}{\sqrt6},
\label{eq:tm1col}
\end{equation}
is the survivor; $\text{TM}_1$ was introduced in Ref.~\cite{XingZhou:2007}, its phenomenology systematized in Ref.~\cite{AlbrightRodejohann:2009}, and it is realized in benchmark form by the littlest seesaw \cite{King:2016}.

The central observation of this paper is that Eq.~(\ref{eq:tm1col}) contains two conditions of different character, and that the data now separate them. The norm condition, $|U_{e1}|^2 = 2/3$, constrains the electron row; it is the first-row correlation $|U_{e1}|^2 = 2(|U_{e2}|^2 + |U_{e3}|^2)$ conjectured in Ref.~\cite{Xing:2026} and is equivalent, given unitarity, to the correlation $\stw = (1-3\sth)/(3\cos^2\theta_{13}) = (1 - 2\tan^2\theta_{13})/3$ tested by JUNO. The balance condition, $|U_{\mu1}| = |U_{\tau1}|$, constrains the $\mu$--$\tau$ structure and is equivalent to a correlation between $\theta_{23}$ and $\cos\delta$. We show in Sec.~\ref{sec:eps} that the norm condition is consistent with current data, deviating by only $1.4\sigma$, while the balance condition fails substantially in central value, $\eps{2} = +0.29$ against zero, yet is disfavored at only $1.5\sigma$ on the profile likelihood, because the phase that drives it remains poorly measured; the first column is trimaximal in norm, and whether it is unbalanced is the question the coming phase measurements settle.

This asymmetry has a natural interpretation. A charged-lepton rotation in the 2--3 sector cannot reach the electron row; it therefore breaks the balance condition while preserving the norm condition, together with every other electron-row prediction, exactly. Sections~\ref{sec:model} and \ref{sec:origin} develop this into a definite model: a $\text{TM}_1$ neutrino sector whose reactor angle is fixed in radicals, $\sth = (2-\sqrt3)/12$, dressed by a single charged-lepton rotation whose angle and phase are constrained by the flavor symmetry. The undressed limit is the intersection of $\text{TM}_1$ with the cobimaximal pattern $\theta_{23} = \pi/4$, $\delta = \pm\pi/2$ of Refs.~\cite{Ma:2015,MaRajasekaran:2017}; we refer to the dressed construction throughout as the $\pi/12$ model, after the single angle that fixes its reactor sector. The construction builds on the $S_4 \times C_4 \times C_3 \times C_2$ model of Ref.~\cite{Krishnan:2019} (here $S_4$ is the permutation group of four objects, of order 24, and $C_n$ denotes the cyclic group of order $n$), which realizes the undressed limit, and on the $\text{TM}_1$ ansatz with internal angle $\theta = \pi/12$ proposed in Ref.~\cite{Zhou:2012}, the angle defined at Eq.~(\ref{eq:base}) below. The same closed forms have recently been rederived in a flavor-spin framework \cite{Jourjine:2026}, where the reactor-angle condition appears as the electron-row identity $4|U_{e2}||U_{e3}| + |U_{e1}|^2 = 1$; we derive this identity and exhibit its equivalent forms in Sec.~\ref{sec:erow}.

The result is a framework with four parameter-free predictions in the electron row, all currently satisfied within $1.5\sigma$, one new parameter that replaces the entire $(\theta_{23},\delta)$ sector, and a mass spectrum inherited unchanged from the neutrino sector. Its sharpest statements are the two it can fail: the CP phase is predicted at $\delta = 272^\circ \pm 2^\circ$, $1.7\sigma$ from the present central value on the current profile likelihood and decidable by Hyper-Kamiokande and DUNE within a decade; and the mass sum $\Sigma m_\nu = 65.6\ \text{meV}$ sits at the current cosmological bound.

\section{Two deviation parameters for the first column}
\label{sec:eps}

Define
\begin{equation}
\eps{1} \equiv \tfrac{3}{2}\,|U_{e1}|^2 - 1, \qquad
\eps{2} \equiv 3\left(|U_{\tau1}|^2 - |U_{\mu1}|^2\right),
\label{eq:epsdef}
\end{equation}
so that exact $\text{TM}_1$ is $\eps{1} = \eps{2} = 0$ and column unitarity closes the system:
\begin{equation}
|U_{\mu1}|^2 = \tfrac16 - \tfrac13\eps{1} - \tfrac16\eps{2}, \qquad
|U_{\tau1}|^2 = \tfrac16 - \tfrac13\eps{1} + \tfrac16\eps{2}.
\end{equation}
In the standard parameterization, with $U$ attached to the $W^-$ vertex as usual (the convention is discussed in Sec.~\ref{sec:quark}),
\begin{align}
\eps{1} &= \tfrac{3}{2}\cos^2\theta_{12}\cos^2\theta_{13} - 1,
\label{eq:eps1}\\
\eps{2} &= 6\, s_{12}c_{12}s_{13}\sin 2\theta_{23}
\left(\cos\delta\big|_{\text{TM}_1} - \cos\delta\right),
\label{eq:eps2}
\end{align}
where
\begin{equation}
\cos\delta\big|_{\text{TM}_1}
= -\frac{\cot 2\theta_{23}\,(1 - 5 s_{13}^2)}{2\sqrt2\, s_{13}\sqrt{1 - 3 s_{13}^2}}
\label{eq:cdtm1}
\end{equation}
is the $\text{TM}_1$ phase correlation \cite{Zhang:2025,DingValle:2024}. Equation~(\ref{eq:eps2}) makes the content of $\eps{2}$ explicit: it measures the departure of the measured phase from the value the balance condition assigns to it at the measured $\theta_{23}$, and it vanishes identically when the correlation holds, whatever $\theta_{23}$ does. The parameter $\eps{1}$ is, up to normalization, the first-row correlation $|U_{e1}|^2 = 2(|U_{e2}|^2 + |U_{e3}|^2)$ of Ref.~\cite{Xing:2026}, proposed there as a conjecture supported by the JUNO and Daya Bay data at the $1\sigma$ level. Two properties established in Ref.~\cite{Xing:2026} carry over to $\eps{1}$: in the canonical seesaw the active--sterile mixing rescales the whole first row by a common factor, so the correlation survives first-row non-unitarity, and the reactor disappearance probability is insensitive to that non-unitarity, so what JUNO measures are the angles of the unitary part of $U$ and what it tests is the correlation itself (direct unitarity tests with reactor antineutrinos are discussed in Ref.~\cite{HuangZhou:2025}). The parameter $\eps{2}$ appears to be new.

Evaluating on the NuFIT~6.1 normal-ordering fit \cite{NuFIT61}, which incorporates JUNO ($\stw = 0.3088^{+0.0067}_{-0.0066}$, $\sth = 0.02248^{+0.00055}_{-0.00059}$, $\stt = 0.470^{+0.017}_{-0.014}$, $\delta = 212^{+26}_{-36}$ degrees), Monte Carlo propagation with two-piece normal errors gives
\begin{equation}
\eps{1} = +0.0135^{+0.0097}_{-0.0098}, \qquad
\eps{2} = +0.290^{+0.067}_{-0.129}.
\label{eq:epsmeas}
\end{equation}
The first is consistent with zero at $1.4\sigma$. For the second, the significance follows from the two-dimensional $(\stt, \delta)$ $\Delta\chi^2$ surface of the global fit itself; minimizing $\Delta\chi^2$ along the $\eps{2}=0$ curve of Eq.~(\ref{eq:cdtm1}) gives $\Delta\chi^2 = 2.3$, reached near $(\stt, \delta) = (0.46, 261^\circ)$, so the balance condition is disfavored at $1.5\sigma$; the result is insensitive to varying $\sth$ across its $1\sigma$ range. This is much weaker than the $3.0\sigma$ that a Gaussian propagation of the marginal errors suggests, and the reason is instructive. The likelihood in $\delta$ is strongly non-Gaussian, with a profile that stays shallow between the best fit and $270^\circ$ ($\Delta\chi^2 = 2.7$ at $\delta = 270^\circ$ on the one-dimensional projection), and the $\eps{2}=0$ curve passes through that shallow region. The two conditions therefore separate in central value rather than in present significance; the best-fit point violates the balance condition substantially, but restoring it costs little while the phase remains poorly measured.

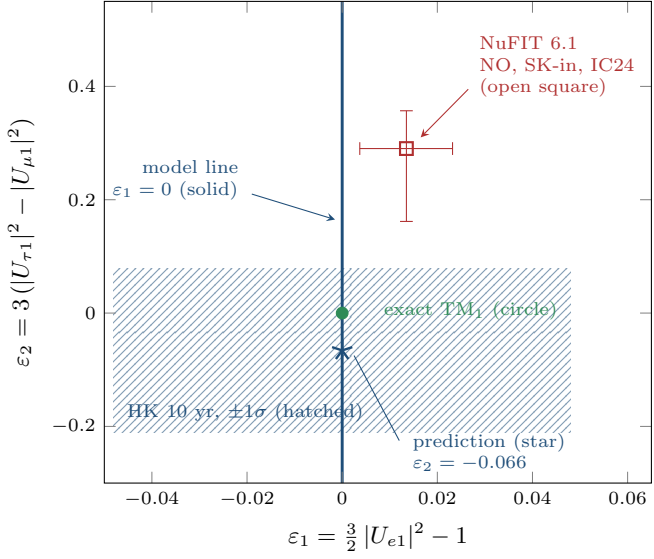
\begin{figure}[t]
\centering
\begin{tikzpicture}
\begin{axis}[width=1.02\columnwidth,height=0.92\columnwidth,
  xmin=-0.05,xmax=0.065,ymin=-0.30,ymax=0.55,
  xlabel={$\varepsilon_1=\tfrac{3}{2}\,|U_{e1}|^2-1$},
  ylabel={$\varepsilon_2=3\,(|U_{\tau1}|^2-|U_{\mu1}|^2)$},
  scaled ticks=false,
  xticklabel style={/pgf/number format/fixed},
  label style={font=\small},tick label style={font=\scriptsize},
  axis on top]
\addplot[pattern=north east lines,pattern color=figblue!55,draw=none,forget plot]
  coordinates {(-0.048,-0.2105) (0.048,-0.2105) (0.048,0.0787) (-0.048,0.0787)} --cycle;
\node[font=\scriptsize,color=figblue,anchor=south west] at (axis cs:-0.047,-0.207)
  {HK 10 yr, $\pm1\sigma$ (hatched)};
\addplot[figblue,line width=1.1pt] coordinates {(0,-0.30) (0,0.55)};
\node[font=\scriptsize,color=figblue,anchor=east,align=right] at (axis cs:-0.020,0.235)
  {model line\\ $\varepsilon_1=0$ (solid)};
\draw[-{stealth},figblue,thin] (axis cs:-0.019,0.21) -- (axis cs:-0.0015,0.165);
\addplot[figred,only marks,mark=square,mark size=2.4pt,line width=0.8pt,forget plot,
  error bars/.cd,x dir=both,x explicit,y dir=both,y explicit]
  coordinates {(0.0135,0.2903) += (0.0097,0.0667) -= (0.0098,0.1286)};
\node[font=\scriptsize,color=figred,anchor=west,align=left] at (axis cs:0.027,0.435)
  {NuFIT 6.1\\ NO, SK-in, IC24\\ (open square)};
\draw[-{stealth},figred,thin] (axis cs:0.0265,0.395) -- (axis cs:0.0165,0.31);
\addplot[figgreen,only marks,mark=*,mark size=2.2pt] coordinates {(0,0)};
\node[font=\scriptsize,color=figgreen,anchor=west] at (axis cs:0.007,0.005) {exact TM$_1$ (circle)};
\addplot[figblue,only marks,mark=star,mark size=4pt,line width=0.9pt] coordinates {(0,-0.0659)};
\node[font=\scriptsize,color=figblue,anchor=west,align=left] at (axis cs:0.013,-0.245)
  {prediction (star)\\ $\varepsilon_2=-0.066$};
\draw[figblue,thin] (axis cs:0.0025,-0.075) -- (axis cs:0.012,-0.225);
\end{axis}
\end{tikzpicture}
\caption{The first-column deviation plane. The model of Sec.~\ref{sec:model} lives on the line $\eps{1}=0$ for any $(\theta_e,\alpha)$; in the real-rotation limit it selects the starred point once $\theta_{23}$ is fixed. The square is the NuFIT~6.1 determination, Eq.~(\ref{eq:epsmeas}), for the normal-ordering (NO) fit that includes Super-Kamiokande atmospheric data and the IceCube 2024 sample (SK-in, IC24: first octant, $\stt = 0.470$, $\delta = 212^\circ$). Of the four NuFIT~6.1 solutions, NO and inverted ordering (IO), each fit with the tabulated Super-Kamiokande atmospheric and IceCube 2024 data (SK-in, IC24) or without them (SK-out, IC23), the NO SK-out solution is nearly identical to the one shown (first octant, $\delta = 207^\circ$); at the second-octant local minimum of the displayed fit ($\Delta\chi^2 = 0.8$: $\stt = 0.550$, $\delta = 185^\circ$) the point would move to $\eps{2} \simeq +0.50$ (Sec.~\ref{sec:eps}); the two IO solutions are not shown. The shaded band is the $\pm1\sigma$ reach of the ten-year Hyper-Kamiokande measurement of $\delta$, centered on the prediction.}
\label{fig:eps}
\end{figure}

The measured $\eps{2}$ carries a caveat worth stating. Its uncertainty is dominated by $\delta$, whose $3\sigma$ range still spans most of the circle, and the second octant survives in the same fit as a local minimum at $\Delta\chi^2 = 0.8$ ($\stt = 0.550$, $\delta = 185^\circ$), where the central value would move to $\eps{2} = +0.50$. The figures display the NO fit with the tabulated Super-Kamiokande atmospheric and IceCube 2024 data (SK-in, IC24) because it is the most complete dataset combination and because normal ordering is the globally preferred one; the NuFIT~6.1 variant without the tabulated atmospheric data (SK-out, IC23) returns a nearly identical first-octant solution ($\stt = 0.470$, $\delta = 207^\circ$), while the second-octant solution of NuFIT~6.0 without SK atmospheric data (SK-out, IC19: $\stt = 0.561$, $\delta = 177^\circ$) is quoted below where it changes a conclusion. Nothing below depends on which variant is adopted; the framework's prediction for $\eps{2}$ is a single number, and any of them tests it. Figure~\ref{fig:eps} displays the situation in the $(\eps{1},\eps{2})$ plane.

\section{The electron row in radicals}
\label{sec:erow}

Given $\text{TM}_1$, one further condition closes the electron row. Reference~\cite{Jourjine:2026} observed that the bilinear
\begin{equation}
Q \equiv 4\,|U_{e2}||U_{e3}| + |U_{e1}|^2
\label{eq:Q}
\end{equation}
equals unity in the data to good accuracy. The identity has a short derivation. Writing $u = |U_{e2}|$, $v = |U_{e3}|$ and using row unitarity $|U_{e1}|^2 = 1 - u^2 - v^2$,
\begin{equation}
Q = 1 + 4uv - u^2 - v^2,
\end{equation}
so that
\begin{equation}
Q = 1 \iff \frac{u}{v} + \frac{v}{u} = 4 \iff \frac{|U_{e2}|}{|U_{e3}|} = 2 + \sqrt3 = \cot\frac{\pi}{12}.
\label{eq:ratio}
\end{equation}
Defining the electron-row angle by $\tan\chi_{13} = |U_{e3}|/|U_{e2}|$, the condition is $\sin 2\chi_{13} = 1/2$, i.e.\ $\chi_{13} = \pi/12$, for any value of $|U_{e1}|$. Combined with the norm condition it fixes the product exactly, $|U_{e2}||U_{e3}| = \tfrac14(1 - |U_{e1}|^2) = 1/12$, equivalently $\sin\theta_{12}\sin 2\theta_{13} = 1/6$, against the measured $0.0824 \pm 0.0014$ ($0.7\sigma$). The identity is therefore independent of the $\text{TM}_1$ norm condition, i.e., of the first-row correlation of Ref.~\cite{Xing:2026}; the two together fix the electron row completely,
\begin{align}
\sth &= \frac{1}{24 + 12\sqrt3} = \frac{2-\sqrt3}{12} = 0.022329,
\label{eq:s13pred}\\
\stw &= \frac{\cos^2(\pi/12)}{2 + \cos^2(\pi/12)} = 0.318107.
\label{eq:s12pred}
\end{align}
These closed forms were first obtained as an ansatz in Ref.~\cite{Zhou:2012} and derived from a flavor group in Ref.~\cite{Krishnan:2019}; Eq.~(\ref{eq:ratio}) shows that the entire content of the reactor-angle condition is the single dimensionless ratio $|U_{e2}|/|U_{e3}| = 2+\sqrt3$. Closed-form proposals for the reactor angle predate these. Reference~\cite{Minkowski:2012} proposed $\theta_{13} = \pi/20$, drawing on the icosahedral group $A_5$, and the tetra-maximal pattern of Ref.~\cite{Xing:2008} gives $\sin\theta_{13} = (\sqrt2-1)/(2\sqrt2)$, $\theta_{13} \simeq 8.4^\circ$, together with $\theta_{23} = \pi/4$ and $\delta = \pi/2$, the same cobimaximal atmospheric sector as the undressed model here, but with $\stw = (10-4\sqrt2)/17 = 0.255$. Against the reactor determination $\sth = 0.02248 \pm 0.00057$, $\pi/20$ gives $0.0245$ ($3.5\sigma$ high), the tetra-maximal value is $0.0214$ ($1.8\sigma$ low), and $\pi/12$ gives $0.0223$ ($0.3\sigma$); the tetra-maximal solar angle is excluded by JUNO. The reactor angle has thus become discriminating among closed forms in its own right.

\begin{figure}[t]
\centering
\begin{tikzpicture}
\begin{axis}[width=1.02\columnwidth,height=0.66\columnwidth,
  xmin=-3.4,xmax=3.4,ymin=-0.6,ymax=3.6,
  xlabel={(measured $-$ predicted) / $\sigma_{\mathrm{meas}}$},
  ytick={0,1,2,3},
  yticklabels={{$Q$},{$|U_{e2}|/|U_{e3}|$},{$\sin^2\theta_{13}$},{$\sin^2\theta_{12}$}},
  title={electron-row predictions (parameter-free)},
  title style={font=\small,yshift=-3pt},
  label style={font=\small},tick label style={font=\scriptsize},
  axis on top]
\addplot[fill=black!4,draw=none,forget plot] coordinates {(-2,-0.6) (2,-0.6) (2,3.6) (-2,3.6)} --cycle;
\addplot[fill=black!9,draw=none,forget plot] coordinates {(-1,-0.6) (1,-0.6) (1,3.6) (-1,3.6)} --cycle;
\addplot[black!55,dashed,thin,forget plot] coordinates {(-2,-0.6) (-2,3.6)};
\addplot[black!55,dashed,thin,forget plot] coordinates {(2,-0.6) (2,3.6)};
\addplot[black!55,dotted,thin,forget plot] coordinates {(-1,-0.6) (-1,3.6)};
\addplot[black!55,dotted,thin,forget plot] coordinates {(1,-0.6) (1,3.6)};
\addplot[figgreen,line width=1.0pt,forget plot] coordinates {(0,-0.6) (0,3.6)};
\addplot[figblue,only marks,mark=*,mark size=2.2pt,line width=0.8pt,
  error bars/.cd,x dir=both,x fixed=1]
  coordinates {(-1.39,3) (0.26,2) (-1.07,1) (1.09,0)};
\end{axis}
\end{tikzpicture}
\caption{Pulls of the four electron-row observables against the parameter-free predictions of Eqs.~(\ref{eq:ratio})--(\ref{eq:s12pred}), in units of the current measurement error; $Q$ is the bilinear of Eq.~(\ref{eq:Q}). The shaded bands mark the $\pm1\sigma$ (darker, dotted edges) and $\pm2\sigma$ (lighter, dashed edges) regions. At the JUNO endgame precision of $\pm0.003$ on $\stw$, the present central value would place the first row at $-3.1\sigma$; that projection is discussed in Sec.~\ref{sec:tests}.}
\label{fig:erow}
\end{figure}
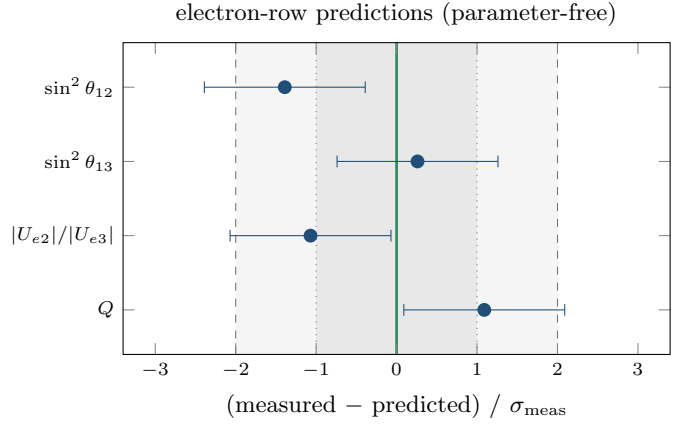

Against NuFIT~6.1,
\begin{align}
\frac{|U_{e2}|}{|U_{e3}|} &= 3.666^{+0.063}_{-0.061}
&&(-1.1\sigma),\nonumber\\
\sth &= 0.02248 \pm 0.00057 &&(+0.3\sigma),\nonumber\\
\stw &= 0.3088 \pm 0.0067 &&(-1.4\sigma),\nonumber\\
Q &= 1.0051^{+0.0046}_{-0.0048} &&(+1.1\sigma),
\label{eq:erowdata}
\end{align}
where the pulls are (measured $-$ predicted)$/\sigma_{\rm meas}$, as in Fig.~\ref{fig:erow}. The four rows of Eq.~(\ref{eq:erowdata}) are not independent tests; they are equivalent statements of two conditions, the first-row correlation of Ref.~\cite{Xing:2026} and the ratio of Eq.~(\ref{eq:ratio}), and we list them because different experiments constrain different combinations. JUNO drives the third line; Daya Bay and the reactor sector drive the second; the first and fourth mix them. All sit within $1.5\sigma$; Fig.~\ref{fig:erow} summarizes the pulls.

\section{The $\pi/12$ model}
\label{sec:model}

\subsection{Structure and invariance}

Let the neutrino sector produce exact $\text{TM}_1$ mixing at the Krishnan point,
\begin{equation}
U_\nu = U_{\text{TM}_1}\!\left(\theta = -\tfrac{\pi}{12},\ \zeta = \tfrac{\pi}{2}\right),
\label{eq:base}
\end{equation}
in the notation of Ref.~\cite{Krishnan:2019}: $\theta$ is the rotation angle of the $\text{TM}_1$ parameterization, acting in the 2--3 block of the neutrino diagonalization with $\sin\theta_{13} = |\sin\theta|/\sqrt3$, and $\zeta$ is its phase. This point yields Eqs.~(\ref{eq:s13pred}) and (\ref{eq:s12pred}) together with $\stt = 1/2$ and $\delta = -\pi/2$; the value $\zeta = \pi/2$ is $\mu$--$\tau$ reflection symmetry \cite{HarrisonScott:2002} and is what enforces $\eps{2} = 0$ in the undressed model. The atmospheric sector of the undressed model is thus the cobimaximal pattern, $\theta_{23} = \pi/4$ with $\delta = \pm\pi/2$ and $\theta_{12}$, $\theta_{13}$ free, named and developed in Refs.~\cite{Ma:2015,MaRajasekaran:2017}; it arises here exactly as it does there, from a trimaximal charged-lepton factor $U_\omega$ acting on a real orthogonal neutrino diagonalization. Dress it with a charged-lepton 2--3 rotation,
\begin{equation}
U = R_{23}(\theta_e,\alpha)^\dagger\, U_\nu,
\qquad
R_{23} = \begin{pmatrix} 1 & 0 & 0\\ 0 & c_e & s_e e^{-i\alpha}\\ 0 & -s_e e^{i\alpha} & c_e \end{pmatrix}.
\label{eq:dressed}
\end{equation}
Because $R_{23}$ acts trivially on the first row, every electron-row quantity is exactly invariant: $\stw$, $\sth$, $Q$, the ratio of Eq.~(\ref{eq:ratio}), $\eps{1} = 0$, and the effective Majorana mass $m_{\beta\beta}$, for all $(\theta_e, \alpha)$. We have verified this numerically to machine precision across the full parameter range. For the first-column balance, with the real column $(\sqrt{2/3}, -1/\sqrt6, -1/\sqrt6)$ one finds the exact relation
\begin{equation}
\eps{2} = \sin 2\theta_e \cos\alpha.
\label{eq:eps2model}
\end{equation}
The single rotation therefore breaks the one condition the data reject and nothing else. The entire $(\theta_{23}, \delta)$ sector becomes a function of $(\theta_e, \alpha)$ on top of the fixed base point.

\subsection{The real-rotation limit}

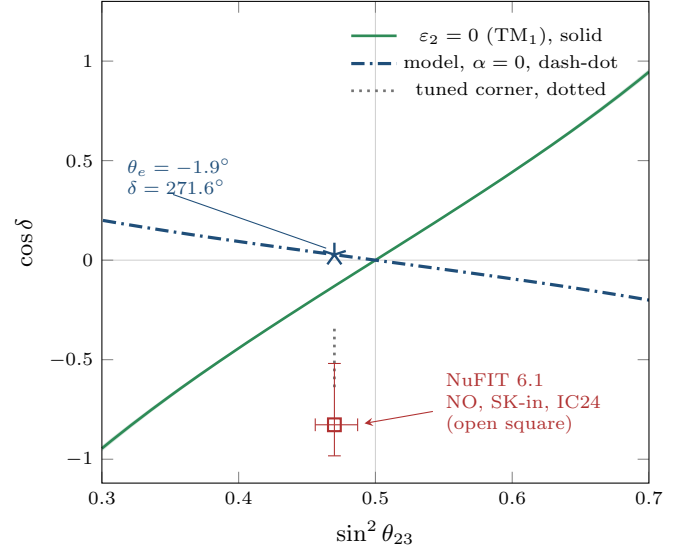
\begin{figure}[t]
\centering
\begin{tikzpicture}
\begin{axis}[width=1.02\columnwidth,height=0.92\columnwidth,
  xmin=0.30,xmax=0.70,ymin=-1.12,ymax=1.30,
  xlabel={$\sin^2\theta_{23}$},ylabel={$\cos\delta$},
  label style={font=\small},tick label style={font=\scriptsize},
  legend style={draw=none,fill=none,font=\scriptsize,at={(0.97,0.97)},anchor=north east},
  axis on top]
\addplot[black!18,thin,forget plot] coordinates {(0.30,0) (0.70,0)};
\addplot[black!18,thin,forget plot] coordinates {(0.5,-1.12) (0.5,1.30)};
\addplot[fill=figgreen!22,draw=none,forget plot] coordinates {(0.3000,-0.9320) (0.3071,-0.8928) (0.3143,-0.8543) (0.3214,-0.8165) (0.3286,-0.7794) (0.3357,-0.7429) (0.3429,-0.7070) (0.3500,-0.6716) (0.3571,-0.6367) (0.3643,-0.6022) (0.3714,-0.5682) (0.3786,-0.5346) (0.3857,-0.5014) (0.3929,-0.4685) (0.4000,-0.4359) (0.4071,-0.4036) (0.4143,-0.3716) (0.4214,-0.3398) (0.4286,-0.3082) (0.4357,-0.2769) (0.4429,-0.2457) (0.4500,-0.2146) (0.4571,-0.1837) (0.4643,-0.1529) (0.4714,-0.1222) (0.4786,-0.0916) (0.4857,-0.0610) (0.4929,-0.0305) (0.5000,-0.0000) (0.5071,0.0314) (0.5143,0.0629) (0.5214,0.0944) (0.5286,0.1259) (0.5357,0.1576) (0.5429,0.1893) (0.5500,0.2212) (0.5571,0.2531) (0.5643,0.2853) (0.5714,0.3176) (0.5786,0.3501) (0.5857,0.3829) (0.5929,0.4159) (0.6000,0.4492) (0.6071,0.4827) (0.6143,0.5166) (0.6214,0.5509) (0.6286,0.5855) (0.6357,0.6206) (0.6429,0.6560) (0.6500,0.6920) (0.6571,0.7285) (0.6643,0.7655) (0.6714,0.8031) (0.6786,0.8414) (0.6857,0.8803) (0.6929,0.9199) (0.7000,0.9603) (0.7000,0.9320) (0.6929,0.8928) (0.6857,0.8543) (0.6786,0.8165) (0.6714,0.7794) (0.6643,0.7429) (0.6571,0.7070) (0.6500,0.6716) (0.6429,0.6367) (0.6357,0.6022) (0.6286,0.5682) (0.6214,0.5346) (0.6143,0.5014) (0.6071,0.4685) (0.6000,0.4359) (0.5929,0.4036) (0.5857,0.3716) (0.5786,0.3398) (0.5714,0.3082) (0.5643,0.2769) (0.5571,0.2457) (0.5500,0.2146) (0.5429,0.1837) (0.5357,0.1529) (0.5286,0.1222) (0.5214,0.0916) (0.5143,0.0610) (0.5071,0.0305) (0.5000,-0.0000) (0.4929,-0.0314) (0.4857,-0.0629) (0.4786,-0.0944) (0.4714,-0.1259) (0.4643,-0.1576) (0.4571,-0.1893) (0.4500,-0.2212) (0.4429,-0.2531) (0.4357,-0.2853) (0.4286,-0.3176) (0.4214,-0.3501) (0.4143,-0.3829) (0.4071,-0.4159) (0.4000,-0.4492) (0.3929,-0.4827) (0.3857,-0.5166) (0.3786,-0.5509) (0.3714,-0.5855) (0.3643,-0.6206) (0.3571,-0.6560) (0.3500,-0.6920) (0.3429,-0.7285) (0.3357,-0.7655) (0.3286,-0.8031) (0.3214,-0.8414) (0.3143,-0.8803) (0.3071,-0.9199) (0.3000,-0.9603)} --cycle;
\addplot[figgreen,line width=1.0pt] coordinates {(0.3000,-0.9459) (0.3071,-0.9061) (0.3143,-0.8671) (0.3214,-0.8287) (0.3286,-0.7911) (0.3357,-0.7540) (0.3429,-0.7175) (0.3500,-0.6816) (0.3571,-0.6462) (0.3643,-0.6112) (0.3714,-0.5767) (0.3786,-0.5426) (0.3857,-0.5089) (0.3929,-0.4755) (0.4000,-0.4424) (0.4071,-0.4096) (0.4143,-0.3771) (0.4214,-0.3449) (0.4286,-0.3128) (0.4357,-0.2810) (0.4429,-0.2493) (0.4500,-0.2178) (0.4571,-0.1865) (0.4643,-0.1552) (0.4714,-0.1241) (0.4786,-0.0930) (0.4857,-0.0620) (0.4929,-0.0310) (0.5000,-0.0000) (0.5071,0.0310) (0.5143,0.0620) (0.5214,0.0930) (0.5286,0.1241) (0.5357,0.1552) (0.5429,0.1865) (0.5500,0.2178) (0.5571,0.2493) (0.5643,0.2810) (0.5714,0.3128) (0.5786,0.3449) (0.5857,0.3771) (0.5929,0.4096) (0.6000,0.4424) (0.6071,0.4755) (0.6143,0.5089) (0.6214,0.5426) (0.6286,0.5767) (0.6357,0.6112) (0.6429,0.6462) (0.6500,0.6816) (0.6571,0.7175) (0.6643,0.7540) (0.6714,0.7911) (0.6786,0.8287) (0.6857,0.8671) (0.6929,0.9061) (0.7000,0.9459)};
\addlegendentry{$\varepsilon_2=0$ (TM$_1$), solid}
\addplot[figblue,line width=1.2pt,dash pattern=on 5pt off 2pt on 1pt off 2pt] coordinates {(0.2435,0.2747) (0.2510,0.2639) (0.2587,0.2533) (0.2664,0.2428) (0.2743,0.2325) (0.2822,0.2224) (0.2902,0.2124) (0.2983,0.2026) (0.3065,0.1929) (0.3147,0.1833) (0.3231,0.1739) (0.3315,0.1645) (0.3400,0.1553) (0.3485,0.1461) (0.3571,0.1371) (0.3657,0.1281) (0.3744,0.1192) (0.3832,0.1104) (0.3920,0.1017) (0.4009,0.0930) (0.4097,0.0843) (0.4187,0.0758) (0.4276,0.0672) (0.4366,0.0587) (0.4456,0.0503) (0.4546,0.0419) (0.4637,0.0335) (0.4728,0.0251) (0.4818,0.0167) (0.4909,0.0084) (0.5000,-0.0000) (0.5091,-0.0084) (0.5182,-0.0167) (0.5272,-0.0251) (0.5363,-0.0335) (0.5454,-0.0419) (0.5544,-0.0503) (0.5634,-0.0587) (0.5724,-0.0672) (0.5813,-0.0758) (0.5903,-0.0843) (0.5991,-0.0930) (0.6080,-0.1017) (0.6168,-0.1104) (0.6256,-0.1192) (0.6343,-0.1281) (0.6429,-0.1371) (0.6515,-0.1461) (0.6600,-0.1553) (0.6685,-0.1645) (0.6769,-0.1739) (0.6853,-0.1833) (0.6935,-0.1929) (0.7017,-0.2026) (0.7098,-0.2124) (0.7178,-0.2224) (0.7257,-0.2325) (0.7336,-0.2428) (0.7413,-0.2533) (0.7490,-0.2639) (0.7565,-0.2747)};
\addlegendentry{model, $\alpha=0$, dash-dot}
\addplot[black!55,dotted,line width=1.0pt] coordinates {(0.4700,-0.6381) (0.4700,-0.5989) (0.4700,-0.5642) (0.4700,-0.5332) (0.4700,-0.5055) (0.4700,-0.4804) (0.4700,-0.4577) (0.4700,-0.4370) (0.4700,-0.4181) (0.4700,-0.4007) (0.4700,-0.3847) (0.4700,-0.3698) (0.4700,-0.3561) (0.4700,-0.3433) (0.4700,-0.3314) (0.4700,-0.3202)};
\addlegendentry{tuned corner, dotted}
\addplot[figred,only marks,mark=square,mark size=2.4pt,line width=0.8pt,forget plot,
  error bars/.cd,x dir=both,x explicit,y dir=both,y explicit]
  coordinates {(0.470,-0.827) += (0.017,0.308) -= (0.014,0.156)};
\node[font=\scriptsize,color=figred,anchor=west,align=left] at (axis cs:0.545,-0.72)
  {NuFIT 6.1\\ NO, SK-in, IC24\\ (open square)};
\draw[-{stealth},figred,thin] (axis cs:0.542,-0.76) -- (axis cs:0.492,-0.82);
\addplot[figblue,only marks,mark=star,mark size=4.5pt,line width=0.9pt] coordinates {(0.4700,0.0276)};
\node[font=\scriptsize,color=figblue,anchor=west,align=left] at (axis cs:0.312,0.42)
  {$\theta_e=-1.9^\circ$\\ $\delta=271.6^\circ$};
\draw[figblue,thin] (axis cs:0.352,0.33) -- (axis cs:0.464,0.06);
\end{axis}
\end{tikzpicture}
\caption{The $(\stt,\cos\delta)$ plane. The rising curve is the undressed $\text{TM}_1$ correlation, $\eps{2}=0$, Eq.~(\ref{eq:cdtm1}); the nearly flat curve is the dressed model in the real-rotation limit, traced by $\theta_e$, along which the phase stays near-maximal while $\theta_{23}$ moves freely. The star marks the prediction at the measured $\stt$. The short dashed segment is the tuned corner ($\theta_e\simeq-29^\circ$, $\alpha$ within a $1.6^\circ$ window) that reproduces the current central value of $\delta$ and is not pursued. The square is the NuFIT~6.1 determination for the NO SK-in IC24 solution, as in Fig.~\ref{fig:eps}.}
\label{fig:phase}
\end{figure}

If the rotation is real, $\alpha = 0$, one parameter remains and the model traces a curve in the $(\stt, \delta)$ plane. Numerically,
\begin{center}
\begin{tabular}{ccccccc}
$\theta_e$ & $-8^\circ$ & $-4^\circ$ & $-2^\circ$ & $0$ & $+4^\circ$ & $+8^\circ$\\
\hline
$\stt$ & 0.375 & 0.437 & 0.468 & 0.500 & 0.563 & 0.625\\
$\delta$ & $277^\circ$ & $273^\circ$ & $272^\circ$ & $270^\circ$ & $267^\circ$ & $263^\circ$\\
$\eps{2}$ & $-0.28$ & $-0.14$ & $-0.07$ & $0$ & $+0.14$ & $+0.28$
\end{tabular}
\end{center}
These numbers follow from closed forms (Appendix~\ref{app:matrix}):
\begin{equation}
\stt = \frac12 + \frac{2+5\sqrt3}{2(10+\sqrt3)}\sin 2\theta_e, \qquad
J = -\frac{\cos 2\theta_e}{12\sqrt6}.
\label{eq:closed23}
\end{equation}
The atmospheric angle responds linearly to the rotation while the Jarlskog invariant is protected quadratically, $J = -\tfrac{1}{12\sqrt6}(1 - 2\theta_e^2 + \cdots)$; this is the analytic content of the flat model curve in Fig.~\ref{fig:phase}, and it is why the phase stays within about $\pm 11^\circ$ of $270^\circ$ over the entire displayed range while $\theta_{23}$ moves freely. Matching the measured $\stt = 0.470$ requires
\begin{equation}
\theta_e = -1.89^\circ,
\qquad\Rightarrow\qquad
\delta = 271.6^\circ,\quad \eps{2} = -0.066,
\label{eq:prediction}
\end{equation}
and $\delta$ varies only between $271^\circ$ and $272.5^\circ$ across the $1\sigma$ range of $\theta_{23}$. Equation~(\ref{eq:prediction}) is the prediction of this paper: near-maximal CP violation, with a small negative first-column asymmetry, at a rotation angle of order the model's own expansion parameter. Figure~\ref{fig:phase} shows the model curve against the $\text{TM}_1$ correlation and the data.

\subsection{Confrontation with data}

We state the present tension plainly, first as a fit. Treating $\theta_e$ as the single free parameter (equivalently the pair $p = \cos\theta_e - \sin\theta_e$, $q = \cos\theta_e + \sin\theta_e$ of Eq.~(\ref{eq:Udressed}), constrained by $p^2 + q^2 = 2$) and minimizing the total $\Delta\chi^2$ assembled from the NuFIT~6.1 NO SK-in IC24 tables, the one-dimensional projections in $\stw$ and $\sth$ evaluated at the fixed electron-row predictions plus the full two-dimensional $(\stt, \delta)$ surface, gives
\begin{equation}
\theta_e = -2.23^{+0.88\circ}_{-0.74}, \;\;
p = 1.0382^{+0.0123}_{-0.0148}, \;\;
q = 0.9603^{+0.0158}_{-0.0135},
\label{eq:fitpq}
\end{equation}
with $\chi^2 = 4.72$ for 3 degrees of freedom (goodness of fit $19\%$). The decomposition locates the tension: the electron row contributes $\Delta\chi^2 = 1.88$ on $\stw$ (a $+1.4\sigma$ pull) and $0.06$ on $\sth$ ($-0.2\sigma$), while the joint $(\stt, \delta)$ surface contributes $2.78$, essentially all of it the phase; at the fitted point the one-dimensional profile in $\stt$ alone costs $0.13$ and that in $\delta$ costs $2.8$. The single parameter absorbs the atmospheric angle, and the electron row stands as in Fig.~\ref{fig:erow}. What remains is the phase: the predicted $\delta = 271.6^\circ$ sits $1.7\sigma$ above the NuFIT~6.1 central value $212^{+26}_{-36}$ on the current profile likelihood, and the predicted $\eps{2} = -0.066$ has the opposite sign to the measured central value in Eq.~(\ref{eq:epsmeas}); the mildness of both numbers reflects the shallow $\delta$ profile of Sec.~\ref{sec:eps}, not agreement in central value.

One aside on the fitted value itself is worth recording. The identification $\sin\theta_e = -|V_{us}|^{9/4} = -0.0349$, equal to $(14/75)^2$ to four decimals and corresponding to $\theta_e = -2.00^\circ$, lies $0.3\sigma$ from Eq.~(\ref{eq:fitpq}). We attach no significance to it absent a mechanism relating the charged-lepton 2--3 rotation to the Cabibbo angle. It is testable in principle: it shifts the prediction to $\stt = 0.4683$, which differs from the unconstrained fit value $0.4647$ by $0.004$, at the edge of the projected DUNE and Hyper-Kamiokande precision, so a sharpened $\theta_{23}$ measurement could eventually distinguish the two.

The fit also locks the octant to the phase. Because $\eps{2} = \sin 2\theta_e$ in this limit, a negative $\theta_e$ pushes $\theta_{23}$ into the first octant and, by the quadratic protection of Eq.~(\ref{eq:closed23}), holds $\delta$ just above $270^\circ$; a positive $\theta_e$ pushes $\theta_{23}$ into the second octant with $\delta$ just below $270^\circ$. Mixed pairings are not available. The second-octant solution of NuFIT~6.0 without SK atmospheric data (SK-out IC19: $\stt = 0.561$, $\delta = 177^\circ$) is such a pairing, and it persists in NuFIT~6.1 as the local minimum at $\Delta\chi^2 = 0.8$ noted in Sec.~\ref{sec:eps}; fitting the NuFIT~6.0 solution with two-piece marginal errors returns $\chi^2 = 23.6$ for 3 degrees of freedom; if that solution is confirmed, the framework is excluded at more than $4\sigma$. A resolved octant therefore tests the construction even before a precision measurement of the phase.

Reproducing the current central value $\delta = 212^\circ$ within Eq.~(\ref{eq:dressed}) is possible but only barely: it requires $\theta_e \simeq -29^\circ$ with $\alpha$ tuned inside a window of roughly $1.6^\circ$ out of $360^\circ$. We regard that corner as a fit rather than a model and do not pursue it. The framework thus stands or falls with the phase: if $\delta$ settles near the current central value, the real-rotation limit is excluded, and with it the simplest version of the construction. Table~\ref{tab:angles} collects the angle and phase definitions of the model and their values.

\begin{table*}[t]
\caption{Angles and phases of the $\pi/12$ model: definitions, what fixes each, and the model value. Measured inputs carry errors; all other values are fixed by the construction or predicted. PMNS angles and phases are given in the standard parameterization, with $U$ attached to the $W^-$ vertex (Sec.~\ref{sec:quark}). The quark-sector phase is listed for the parallel of Sec.~\ref{sec:quark}.}
\label{tab:angles}
\begin{ruledtabular}
\begin{tabular}{llll}
Symbol & Definition & Fixed by & Value \\
\hline
$\theta$ & $\text{TM}_1$ rotation angle, $\sin\theta_{13} = |\sin\theta|/\sqrt3$, Eq.~(\ref{eq:base}) & $\langle\phi\rangle$ alignment & $-\pi/12 = -15^\circ$ \\
$\zeta$ & $\text{TM}_1$ phase, Eq.~(\ref{eq:base}) & $\mu$--$\tau$ reflection (residual CP) & $\pi/2$ \\
$\alpha_\phi$ & alignment director angle of $\langle\phi\rangle$, $\theta = \alpha_\phi/2$ (Sec.~\ref{sec:origin}) & $Y_{24}$ residual symmetries & $-\pi/6$ (two ticks of $e^{i\pi/12}$) \\
$\chi_{13}$ & electron-row angle, $\tan\chi_{13} = |U_{e3}|/|U_{e2}|$, Eq.~(\ref{eq:ratio}) & $= |\theta|$; parameter-free prediction & $\pi/12 = 15^\circ$ \\
$\theta_e$ & charged-lepton 2--3 rotation angle, Eq.~(\ref{eq:dressed}) & fitted (the single free parameter) & $-2.23^{+0.88}_{-0.74}{}^\circ$ \\
$\alpha$ & phase of the charged-lepton rotation, Eq.~(\ref{eq:dressed}) & $\xi$ alignment (Appendix~\ref{app:flavon}) & $0$ \\
$\theta_{12}$ & PMNS solar angle & predicted, Eq.~(\ref{eq:s12pred}) & $34.33^\circ$ ($\stw = 0.31811$) \\
$\theta_{13}$ & PMNS reactor angle & predicted, Eq.~(\ref{eq:s13pred}) & $8.59^\circ$ ($\sth = 0.02233$) \\
$\theta_{23}$ & PMNS atmospheric angle & linear in $\theta_e$, Eq.~(\ref{eq:closed23}) & $42.97^\circ$ ($\stt = 0.46467$) \\
$\delta$ & PMNS Dirac phase & quadratically protected, Eq.~(\ref{eq:closed23}) & $271.9^\circ$ \\
$(\alpha_{21}, \alpha_{31})$ & Majorana phases, PDG convention, Eq.~(\ref{eq:finalangles}) & sign pattern $\eta = (+,+,-)$ & $(0,\ \pi)$ \\
$\phi_{\rm FX}$ & Fritzsch--Xing quark phase, Sec.~\ref{sec:quark} & extracted from CKM moduli & $92.7^\circ \pm 3.7^\circ$ \\
\end{tabular}
\end{ruledtabular}
\end{table*}

\section{Origin of the rotation}
\label{sec:origin}

The undressed base point, Eq.~(\ref{eq:base}), is realized by the $S_4 \times C_4 \times C_3 \times C_2$ model of Ref.~\cite{Krishnan:2019}: the charged-lepton mass matrix is left-diagonalized exactly by the trimaximal matrix $U_\omega$, independent of the Yukawa couplings, because the flavon $\chi$ \cite{FN:1979} with $\langle\chi\rangle \propto (1,\omega,\bar\omega)$, $\omega \equiv e^{2\pi i/3}$, supplies all three columns; the neutrino Dirac matrix is proportional to the identity; and the Majorana triplet alignment $\langle\phi\rangle \propto (-\tfrac{1}{2\sqrt2}, -\sqrt3, \tfrac{1}{2\sqrt2})$ produces a 2--3 block whose mixing angle is half the alignment angle, giving $\theta = -\pi/12$.

The geometric origin of the $\pi/12$ can be stated at three levels, the third displayed in the right panel of Fig.~\ref{fig:z2}. First, $S_4$ is the rotation group of a cube, and the triplet alignments its residual symmetries can define uniquely are the distinguished directions of that cube: the vertices, the face centers, and the edge midpoints, i.e.\ the cube, octahedron, and cuboctahedron orbits identified in Ref.~\cite{Krishnan:2019}. The $\phi$ direction is none of them; this is the geometric statement of why the $\pi/12$ requires structure beyond $S_4$. Second, in the 2--3 block of the Majorana matrix the traceless perturbation defined by $\langle\phi\rangle$ is a director at angle $\alpha_\phi = -\pi/6$ in the plane of the diagonal difference and the off-diagonal entry (the subscript distinguishing it from the rotation phase $\alpha$ of Sec.~\ref{sec:model}), and the eigenframe of a symmetric matrix turns at half the director angle, giving $\theta = \alpha_\phi/2 = -\pi/12$; both of the model's angles are therefore half-angles, $\theta_{23} = \tfrac12(\pi/2)$ from the reflection and $\theta = \tfrac12(\pi/6)$ from the alignment. Third, the $\pi/6$ itself is quantized: the auxiliary group $Y_{24}$ is generated with the phases $e^{i\pi/12}$ \cite{Krishnan:2019}, so orientation in the enlarged flavor space comes in ticks of $15^\circ$; the residual symmetries fix the elementary alignments, and the invariant product transfers their relative orientation, two ticks, to the effective triplet. The angle $\chi_{13} = \pi/12$ is one tick of the auxiliary clock, as $\delta$ is one quarter-turn of the fourth-root clock of Appendix~\ref{app:matrix}.

Two properties of this construction matter for the extension.

First, the model protects its own alignment, and it does so completely at the level of $\chi$ bilinears: each of $(\chi^*\chi^*)_{\bm{3'}}$, $(\chi\chi)_{\bm{3'}}$, and $(\chi^*\chi)_{\bm{3'}}$ carries exactly the $C_3$ charge that routes it into the column whose alignment it reproduces (the subscript denotes projection onto the triplet representation $\bm{3'}$ of $S_4$; Appendix~\ref{app:flavon}), so higher-order operators built from $\chi$ alone renormalize Yukawa couplings without rotating the left-handed basis. Generating $\theta_e \neq 0$ requires a genuinely new flavon, $\xi$, entering the $\tau_R$ column at relative order $v_\rho/\Lambda$,
\begin{equation}
y_\xi\, \bar L\, \frac{\rho\, \xi}{\Lambda^2}\, \tau_R H .
\end{equation}

Second, the phase $\alpha$ is fixed by the alignment of $\xi$, not free. The induced rotation obeys $\theta_e e^{i\alpha} \propto [U_\omega \langle\xi\rangle]_2 / m_\tau$, where $[X]_i$ denotes the $i$th component of the vector $X$ (not a modulus) and $\langle\xi\rangle$ the vacuum alignment, and evaluating the candidate directions: $(1,1,1)$ and $(1,\omega,\bar\omega)$ give no rotation at all; $(0,1,-1)$ gives a purely imaginary entry, $\alpha = \pi/2$, hence $\eps{2} = 0$ and no observable effect on the balance; and the conjugate-trimaximal direction
\begin{equation}
\langle\xi\rangle \propto (1, \bar\omega, \omega)
\quad\Rightarrow\quad
[U_\omega\langle\xi\rangle]_2 = \sqrt3 \ \ (\text{real}),\ \ \alpha = 0,
\end{equation}
is the unique alignment among these that produces a real rotation, with the same phase structure as $m_\tau$ itself. The real-rotation limit of Sec.~\ref{sec:model} is therefore not an assumption of convenience; it is the statement that $\xi$ takes the $\langle\chi^*\rangle$ alignment, which the $C_3$ charge assignment permits since charge and alignment are independent data. A renormalizable potential that selects this alignment is exhibited in Appendix~\ref{app:flavon}: a Lagrange identity makes one charge-allowed $\xi$--$\chi$ invariant vanish exactly and only at the conjugate-trimaximal direction, which global numerical minimization confirms as the vacuum in an open region of couplings, with $\alpha = 0$ exact at the minimum and with back-reaction on $\langle\phi\rangle$ and $\langle s\rangle$ entering only through operators already present in the base model. The embedding into the $Y_{24}$ construction of Ref.~\cite{Krishnan:2019} is also carried out there: with $\xi$ neutral under the auxiliary group, every allowed coupling of $\xi$ to the auxiliary sector is of a class the recast model already contains with $\chi$, all operators linear in $\xi$ vanish at the vacuum, and direct mixing with the effective triplet is forbidden by $C_3$ and $C_4$; the demonstration is complete at the operator level, with only the joint minimization of the full multi-flavon potential across sectors left to the same sector-by-sector treatment the base model itself employs.

The group-theoretic setting is one realization, not the only one. The undressed structure also arises from residual symmetries: $S_4$ combined with generalized CP, broken to $Z_2 \times Z_2^{\rm CP}$, yields $\text{TM}_1$ mixing with $\delta$ either conserved or maximal \cite{LiDing:2013}, the same intersection realized here by the alignments of Ref.~\cite{Krishnan:2019}; for a comprehensive review of the residual flavour and CP approach see Ref.~\cite{DingValle:2024}. Two differences matter. In the residual-symmetry setting the $\text{TM}_1$ rotation angle, and with it $\theta_{13}$, remains a continuous free parameter, and the comprehensive scan of finite groups with generalized CP in Ref.~\cite{YaoDing:2016} finds fixed columns and phases but no group that promotes $\chi_{13} = \pi/12$ to a symmetry prediction; the alignment origin of the $\pi/12$ is therefore essential rather than convenient. And the residual-symmetry models predict trivial Majorana phases \cite{LiDing:2013}, whereas the present construction forces $\alpha_{31} = \pi$, so $m_{\beta\beta}$ distinguishes the two routes: $5.4$ meV here against $7.7$ meV for trivial phases with the same spectrum.

The size comes out naturally. With the model's expansion parameters $v_\chi/\Lambda \approx v_\rho/\Lambda \approx 0.012$ fixed by the charged-lepton mass hierarchy, the induced angle is $\theta_e \sim (y_\xi/y_\tau)(v_\rho/\Lambda)(v_\xi/v_\chi) \approx 0.7^\circ$ for unit ratios, against the required $1.9^\circ$; an $O(1)$ coefficient near $2.7$ suffices, comparable to the $y_\mu \approx 2.4$ the model already carries.

\section{Masses and external constraints}
\label{sec:masses}

Because the charged-lepton rotation does not act on the neutrino mass matrix, the spectrum of Ref.~\cite{Krishnan:2019} is inherited unchanged. Refitting its two parameters to the current mass-squared differences, we obtain
\begin{equation}
m_1 = 5.2,\quad m_2 = 10.1,\quad m_3 = 50.4\ \text{meV},
\end{equation}
hence
\begin{equation}
\Sigma m_\nu = 65.6\ \text{meV}, \qquad m_{\beta\beta} = 5.4\ \text{meV}.
\label{eq:masspred}
\end{equation}
Because Ref.~\cite{Krishnan:2019} is unpublished, we have verified its construction independently from the Lagrangian: the exact, Yukawa-independent $U_\omega$ diagonalization of the charged-lepton sector, the $\langle\phi\rangle$-alignment eigenvalues $(-\sqrt3,\,\sqrt3/2-1,\,\sqrt3/2+1)$ with mixing angle $-\pi/12$, the resulting $\text{TM}_1$ angles, the $\mu$--$\tau$ reflection, and the two-parameter mass sector all check numerically. Two corrections emerged. The value of $m_2$ quoted there, 8.5--9.3 meV, is inconsistent with its own $m_1$ and $\Delta m^2_{21}$ ranges; the corrected value is near 10.1 meV, with $\Sigma m_\nu$ unaffected. More consequentially, on the branch that fits the mass-squared differences the smallest eigenvalue of $M_{ss}$ is negative, so $\nu_3$ carries a relative Majorana sign, $\eta_3 = -1$. Computing $m_{\beta\beta}$ convention-free as the $ee$ element of $m_\nu$ in the charged-lepton-diagonal basis gives $5.4$ meV; the range $7.2$--$7.7$ meV quoted in Ref.~\cite{Krishnan:2019} is the all-positive combination, which omits this sign.

The mass sum is now the binding external constraint. The DESI DR2 combination with CMB data gives $\Sigma m_\nu < 64.2\ \text{meV}$ at 95\% C.L.\ in $\Lambda$CDM, with a Feldman--Cousins limit of $53\ \text{meV}$ \cite{DESI:2025}; the prediction of Eq.~(\ref{eq:masspred}) sits at or just beyond the bound. The limit relaxes to roughly $160\ \text{meV}$ under CPL dynamical dark energy and to $92\ \text{meV}$ with a hierarchy-informed prior \cite{DESI:2025,Hou:2026}. The framework therefore requires either dynamical dark energy or a modification of its neutrino sector; we state this as a condition of viability rather than a difficulty to be minimized. The prediction $m_{\beta\beta} = 5.4\ \text{meV}$ is below the reach of current $0\nu\beta\beta$ experiments but within the target range of next-generation programs.

These mass predictions also discriminate against the main competing explanation of the $\stw$ offset. Renormalization-group running can reconcile exact $\text{TM}_1$, and even $\text{TM}_2$, with the JUNO measurement, but only for quasi-degenerate neutrino masses \cite{Zhang:2025}, which the spectrum above is not; conversely the present framework needs nothing from the running. For the hierarchical spectrum above ($\Sigma m_\nu = 65.6\ \text{meV}$), renormalization-group corrections to the model's own mixing predictions are negligible. Cosmology and KATRIN therefore separate the two mechanisms now, without waiting for the phase measurement.

\section{Experimental tests}
\label{sec:tests}

The framework makes five statements testable on distinct timescales.

(i) \emph{JUNO endgame.} The final JUNO precision of roughly $\pm 0.003$ on $\stw$ tests Eq.~(\ref{eq:s12pred}) at the $3\sigma$ level if the central value holds at $0.309$. This is the sharpest test of the norm condition and of the entire electron-row sector.

(ii) \emph{The phase.} The prediction $\delta = 272^\circ \pm 2^\circ$ is separated from the current central value $212^\circ$ by $0.88$ in $\cos\delta$, equivalently $0.36$ in $\eps{2}$. Discrimination at $2\sigma$, $3\sigma$, and $5\sigma$ requires $\sigma(\delta) \approx 28^\circ$, $19^\circ$, and $11^\circ$ near $\delta \approx 240^\circ$; these are Gaussian statements, appropriate to a dedicated measurement whose likelihood in $\delta$ is approximately Gaussian at the projected precision, and the shallow profile that limits the present significance in Sec.~\ref{sec:eps} is exactly what such a measurement removes. The Hyper-Kamiokande accelerator program projects $\sigma(\delta) = 20^\circ$ at maximal CP violation with the full ten-year exposure \cite{HK:2025}, reaching the $3\sigma$ separation alone; the combination with DUNE goes beyond it.

(iii) \emph{The octant correlation.} In the real-rotation limit, the sign of $\theta_e$ ties the octant to the phase: $\stt < 1/2$ requires $\delta$ slightly above $270^\circ$, and $\stt > 1/2$ slightly below. A resolved octant plus a measured $\delta$ tests the correlation independently of its overall normalization.

(iv) \emph{Cosmology.} $\Sigma m_\nu = 65.6\ \text{meV}$ is rigid. A $\Lambda$CDM bound securely below $60\ \text{meV}$ excludes the model's neutrino sector outright.

(v) \emph{$0\nu\beta\beta$.} $m_{\beta\beta} = 5.4\ \text{meV}$, unchanged by the extension, with the Majorana sign pattern $\eta = (+,+,-)$ fixed by the construction.

\section{A quark-sector counterpart}
\label{sec:quark}

We close with an observation, presented as a parallel rather than a derivation. In the Fritzsch--Xing parameterization \cite{FX:1997}, $V_{\rm CKM} = R_{12}(\theta_u)R_{23}(\theta)\,\text{diag}(e^{-i\phi},1,1)R_{12}^\dagger(\theta_d)$, the measured moduli \cite{PDG2024} give $\phi = 92.7^\circ \pm 3.7^\circ$, consistent with a maximal phase. As with the lepton sector, the statement is best made convention-free: $\phi = \pi/2$ eliminates the interference between the two 1--2 rotations\footnote{In this parameterization $V_{us} = s_u c_\theta c_d - s_d c_u\, e^{-i\phi}$ exactly, with $s_{u,d} \equiv \sin\theta_{u,d}$, so $|V_{us}|^2 = (s_u c_\theta c_d)^2 + (s_d c_u)^2 - 2\, s_u c_u s_d c_d c_\theta \cos\phi$; the last term is the interference, and $\phi = \pi/2$ removes it. Using the exact identifications $s_u = |V_{ub}|/\sqrt{|V_{ub}|^2 + |V_{cb}|^2}$, $s_d = |V_{td}|/\sqrt{|V_{td}|^2 + |V_{ts}|^2}$, $c_\theta = |V_{tb}|$, together with the identity $|V_{td}|^2 + |V_{ts}|^2 = |V_{ub}|^2 + |V_{cb}|^2$ of this parameterization, the quadrature sum becomes Eq.~(\ref{eq:vus}).} and is equivalent to the magnitude relation
\begin{equation}
|V_{us}|^2 = \frac{(|V_{ub}||V_{ts}||V_{tb}|)^2 + (|V_{cb}||V_{td}|)^2}{(|V_{ub}|^2 + |V_{cb}|^2)^2},
\label{eq:vus}
\end{equation}
which evaluates to $|V_{us}| = 0.2213 \pm 0.0060$ against the measured $0.22501 \pm 0.00068$, a $1.6\%$ statement at $0.6\sigma$. Both sectors are therefore compatible with a structural, near-maximal phase: in the leptons because $\mu$--$\tau$ reflection fixes $\delta = -\pi/2$ up to the small real rotation of Eq.~(\ref{eq:prediction}), in the quarks because the Fritzsch--Xing interference term is compatible with zero. Figure~\ref{fig:rightangle} displays the two phases on a common circle.

Two conventions enter this parallel and should be made explicit. First, mixing angles and phases are parameterization-dependent: $\delta$ is the phase of the standard parameterization and $\phi$ that of the Fritzsch--Xing one, and a phase maximal in one is not maximal in the other (the standard-parameterization phase of the same CKM moduli is $\delta_{\rm CKM} \approx 66^\circ$ \cite{PDG2024}). Each maximality statement therefore stands for the relation among moduli it is equivalent to. For the quarks that relation is Eq.~(\ref{eq:vus}); for the leptons, $\cos\delta = 0$ is the vanishing of the interference term $\pm 2 s_{12}c_{12}s_{13}s_{23}c_{23}\cos\delta$ in $|U_{\mu1}|^2$ and $|U_{\tau1}|^2$, which at $\theta_{23} = \pi/4$ is the $\mu$--$\tau$ balance $|U_{\mu i}| = |U_{\tau i}|$. These relations, and the Jarlskog invariant, are the parameterization-independent content of the comparison; the parallel is that each sector admits a parameterization in which its phase is a right angle, not that the two phases are the same parameter. Second, the two matrices are conventionally attached to opposite charged-current vertices, $V$ to $\bar u_L\gamma^\mu V d_L W^+_\mu$ and $U$ to $\bar\ell_L\gamma^\mu U \nu_L W^-_\mu$, so the lepton counterpart of $V$ at a common vertex is $U^\dagger$, whose entries are the complex conjugates of those of $U$ with rows and columns exchanged. Conjugation reverses the sign of the Jarlskog invariant and of $\sin\delta$ and leaves every modulus and $\cos\delta$ unchanged, so the relative sign of $J_q$ and $J_\ell$ is fixed only once a common vertex is chosen, and in that convention the lepton arrow of Fig.~\ref{fig:rightangle} is reflected through the horizontal axis onto the same side as the quark arrow ($\delta = 272^\circ \to 88^\circ$). All lepton phases in this paper are quoted in the standard $W^-$ convention. The relative orientation of the two right angles is thus a convention. What is not a convention is that each phase lies within its errors of $\pm\pi/2$: $\cos\delta$ and $\cos\phi$ are unchanged by conjugation, and each equals zero exactly when the corresponding relation among moduli holds, Eq.~(\ref{eq:vus}) for the quarks and the vanishing interference term for the leptons.

Two distinct principles could make an interference term vanish, and they are testable against each other. The deterministic one is symmetry: CP is a $Z_2$ transformation, so a residual CP symmetry quantizes a phase to be either conserved or maximal, with nothing in between; this is the mechanism at work in the lepton sector here, where the reality of the Majorana matrix together with the trimaximal charged-lepton factor forces the fourth root of unity in Eq.~(\ref{eq:Unu}) (Appendix~\ref{app:matrix}), and it is the mechanism surveyed in Refs.~\cite{LiDing:2013,YaoDing:2016}. The statistical alternative is incoherence: if an element receives contributions from many sources with independent phases, the cross terms self-average and the moduli add in quadrature. A single randomly drawn phase, by contrast, explains nothing; the probability that one flat phase lands within the measured $3.7^\circ$ of $\pm\pi/2$ is $4\%$, and that two sectors do so independently, $2\times10^{-3}$, so the pattern of Fig.~\ref{fig:rightangle} is either structural or a coincidence at the per-mille level. The two surviving principles separate observably. A residual symmetry fixes the sign of the Jarlskog invariant, as the $\pi/12$ model does with $J_\ell < 0$ in the standard convention, and can correlate the two sectors once a common convention is fixed; incoherence leaves the signs uncorrelated and predicts no relation between them. The construction of this paper sits on the deterministic side of each fork.

\begin{figure}[t]
\centering
\begin{tikzpicture}[scale=1.0]
\def\R{2.55}
% measured delta wedges: 2 sigma (140-264), then 1 sigma (176-238)
\fill[figred!7] (0,0) -- (140:\R) arc (140:264:\R) -- cycle;
\fill[figred!16] (0,0) -- (176:\R) arc (176:238:\R) -- cycle;
% model wedge 271.6 +- 2 (drawn to scale)
\fill[figblue!35] (0,0) -- (269.6:\R) arc (269.6:273.6:\R) -- cycle;
% quark wedges: 2 sigma (85.3-100.1), then 1 sigma (89.0-96.4)
\fill[figgreen!12] (0,0) -- (85.3:\R) arc (85.3:100.1:\R) -- cycle;
\fill[figgreen!28] (0,0) -- (89.0:\R) arc (89.0:96.4:\R) -- cycle;
% circle and axes
\draw[black!70] (0,0) circle (\R);
\draw[black!40,thin] (-\R-0.25,0) -- (\R+0.25,0);
\draw[black!40,thin] (0,-\R-0.25) -- (0,\R+0.25);
% right-angle markers at origin
\draw[black!80,thick] (0.34,0) -- (0.34,-0.34) -- (0,-0.34);
\draw[black!80,thick] (0.34,0) -- (0.34,0.34) -- (0,0.34);
% arrows
\draw[-{stealth},figgreen,line width=1.2pt] (0,0) -- (92.7:\R);
\draw[-{stealth},figblue,line width=1.3pt] (0,0) -- (271.6:\R);
\draw[-{stealth},figred,line width=1.0pt,dashed] (0,0) -- (212:\R);
% degree labels
\node[font=\scriptsize,black!70] at (0:\R+0.42) {$0^\circ$};
\node[font=\scriptsize,black!70] at (90:\R+0.42) {$90^\circ$};
\node[font=\scriptsize,black!70] at (180:\R+0.45) {$180^\circ$};
\node[font=\scriptsize,black!70] at (270:\R+0.42) {$270^\circ$};
% annotations
\node[font=\scriptsize,figgreen,align=center,anchor=south west] at (60:\R+0.10)
  {quarks\\ $\phi_{\rm FX}=92.7^\circ\pm3.7^\circ$};
\node[font=\scriptsize,figblue,align=center,anchor=north west] at (300:\R-0.35)
  {leptons, $\pi/12$ model\\ $\delta=271.6^\circ\pm2^\circ$};
\node[font=\scriptsize,figred,align=center,anchor=north east] at (196:\R-0.12)
  {measured\\ $\delta=212^{+26\circ}_{-36}$};
\end{tikzpicture}
\caption{The two right angles. Phases are drawn as vectors on the unit circle with their $1\sigma$ (dark) and $2\sigma$ (light) wedges, the $2\sigma$ ranges taken as twice the two-piece $1\sigma$ errors. Downward: the lepton Dirac phase of the $\pi/12$ model, pinned near $270^\circ$ by the quadratic protection of Eq.~(\ref{eq:closed23}) (solid arrow, narrow wedge drawn to scale), against the measured NuFIT~6.1 value (dashed arrow, wide wedge). Upward: the quark Fritzsch--Xing phase extracted from the CKM moduli, Sec.~\ref{sec:quark}. The corner marks at the origin are the two exact right angles, $\delta = -\pi/2$ from $\mu$--$\tau$ reflection and $\phi_{\rm FX} = +\pi/2$ from vanishing interference; each sector sits within $1\sigma$ of its right angle except the measured lepton phase, whose resolution is the test of Sec.~\ref{sec:tests}. The mirror orientation reflects opposite signs of the Jarlskog invariants, $J_q > 0$ and $J_\ell < 0$ in the model, in the standard conventions that attach $V$ to the $W^+$ vertex and $U$ to the $W^-$ vertex; referred to a common vertex, the lepton arrow is reflected through the horizontal axis and the two phases lie on the same side of the circle (Sec.~\ref{sec:quark}). No relation between individual CKM and PMNS magnitudes is implied.}
\label{fig:rightangle}
\end{figure}
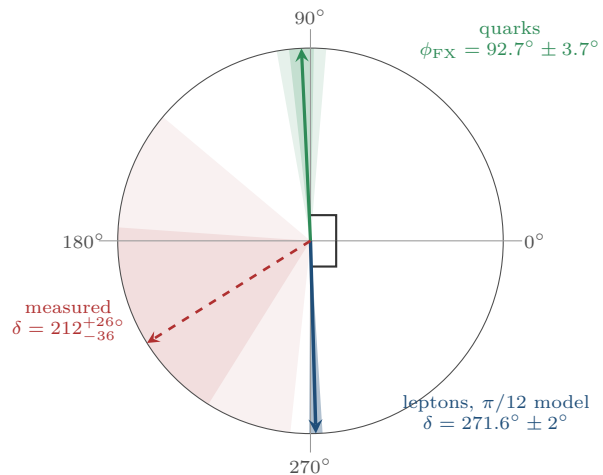 Whether a single symmetry principle enforces both is a question for a companion analysis; here we note only that the lepton-sector prediction of this paper, if confirmed, would make the parallel exact at the few-degree level.

\section{Conclusion}
\label{sec:conc}

The first column of the lepton mixing matrix separates, under current data, into a trimaximal norm that holds and a $\mu$--$\tau$ balance that fails in central value, at a present significance of $1.5\sigma$ set by the still poorly measured phase. A charged-lepton 2--3 rotation acting on a $\text{TM}_1$ neutrino sector is the minimal structure with exactly this signature: it preserves the four electron-row predictions, $\stw = 0.31811$, $\sth = (2-\sqrt3)/12$, $|U_{e2}|/|U_{e3}| = 2+\sqrt3$, and $m_{\beta\beta} = 5.4\ \text{meV}$, all currently satisfied, while generating $\eps{2} = \sin 2\theta_e\cos\alpha$. The structural core admits a one-sentence statement: a single $\mu$--$\tau$ reflection fixes both maximal quantities, the atmospheric angle to half a right angle through its unitary half, the mirror line bisecting the flavor axes, and the Dirac phase to a right angle through its antiunitary half, the fourth roots of unity of Appendix~\ref{app:matrix}; the charged-lepton rotation breaks the first linearly in $\theta_e$ while protecting the second quadratically. The data currently break exactly the fragile half. In the symmetry-motivated real-rotation limit the framework accounts for the departure of $\theta_{23}$ from maximal at a natural rotation angle and predicts that the surviving right angle stands: near-maximal CP violation, $\delta = 272^\circ \pm 2^\circ$, in tension with the present central value at $1.7\sigma$ on the current profile likelihood and decidable by Hyper-Kamiokande and DUNE. The rigid mass sum $\Sigma m_\nu = 65.6\ \text{meV}$ sits at the current DESI bound and requires dynamical dark energy if $\Lambda$CDM tightens further. Either the phase measurement or cosmology can refute the $\pi/12$ model within the decade; the electron row will have been tested by JUNO on the same timescale. Few frameworks of comparable economy are this completely falsifiable, and that falsifiability, rather than the present quality of fit, is the argument for taking it seriously.

\begin{acknowledgments}
V.B. thanks Alex Jourjine for stimulating correspondence on this topic.
V.B. gratefully acknowledges support from the U.S. Department of Energy, Office of Science, Office of High Energy Physics, under Award Number DE-SC0017647 and from the William F. Vilas Estate.
\end{acknowledgments}

\vspace{4pt}
\begin{center}\rule{0.55\columnwidth}{0.6pt}\end{center}
\vspace{2pt}

\appendix

\section{The mixing matrix in closed form}
\label{app:matrix}

\emph{The undressed matrix.} Write $s_\pi \equiv \sin(\pi/12) = (\sqrt6-\sqrt2)/4$ and $c_\pi \equiv \cos(\pi/12) = (\sqrt6+\sqrt2)/4$. The undressed matrix of Eq.~(\ref{eq:base}) is, up to unphysical phases,
\begin{widetext}
\begin{equation}
\setlength{\arraycolsep}{9pt}
U_\nu =
\begin{pmatrix}
\sqrt{\dfrac{2}{3}} & \dfrac{c_\pi}{\sqrt3} & i\,\dfrac{s_\pi}{\sqrt3}\\[8pt]
-\dfrac{1}{\sqrt6} & \dfrac{c_\pi}{\sqrt3} + i\,\dfrac{s_\pi}{\sqrt2} & \dfrac{c_\pi}{\sqrt2} + i\,\dfrac{s_\pi}{\sqrt3}\\[8pt]
-\dfrac{1}{\sqrt6} & \dfrac{c_\pi}{\sqrt3} - i\,\dfrac{s_\pi}{\sqrt2} & -\dfrac{c_\pi}{\sqrt2} + i\,\dfrac{s_\pi}{\sqrt3}
\end{pmatrix},
\label{eq:Unu}
\end{equation}
in which all four standard parameters are fixed: $\theta_{12}$ and $\theta_{13}$ by Eqs.~(\ref{eq:s13pred})--(\ref{eq:s12pred}), $\theta_{23} = 45^\circ$, and $\delta = 270^\circ$. The electron row reads off directly as $(\sqrt{2/3},\ c_\pi/\sqrt3,\ s_\pi/\sqrt3)$; both $\sin\theta_{13} = \sin(\pi/12)/\sqrt3$ and the ratio $|U_{e2}|/|U_{e3}| = \cot(\pi/12)$ of Eq.~(\ref{eq:ratio}) are visible at sight. The $\mu$ and $\tau$ rows are complex conjugates up to the sign of the third column; this is the $\mu$--$\tau$ reflection.

The purely imaginary entry is forced by the reflection. The reflection acts antiunitarily: it exchanges the $\mu$ and $\tau$ rows and conjugates. Applying it twice gives the identity, so it can at most multiply each column by a sign, $\kappa_i = \pm1$ (distinct from the Majorana signs $\eta_i$ of Sec.~\ref{sec:tests}). The electron-row entries then obey $U_{ei}^* = \kappa_i U_{ei}$, and the fourth root of unity follows in two lines: writing $U_{ei} = |U_{ei}|\, e^{i\varphi_i}$, the condition reads $e^{-i\varphi_i} = \kappa_i\, e^{i\varphi_i}$, i.e.\ $e^{2i\varphi_i} = \kappa_i$, hence $e^{4i\varphi_i} = \kappa_i^2 = 1$ and $\varphi_i \in \{0,\ \pi/2,\ \pi,\ 3\pi/2\}$. For $\kappa_i = +1$ the entry is real ($e^{2i\varphi_i} = 1$); for $\kappa_i = -1$ it is purely imaginary ($e^{i\varphi_i} = \pm i$). The third column carries $\kappa_3 = -1$. That sign is the $i$ in Eq.~(\ref{eq:Unu}), and in the standard parameterization it is maximal Dirac CP violation: $\delta = \pm\pi/2$, with no intermediate value available.

The same reflection fixes $\theta_{23} = \pi/4$, and the reason is geometric. The mirror line of the $\mu$--$\tau$ exchange bisects the right angle between the flavor axes, and a mass eigenstate mapped to itself is pinned to that bisector, $(\hat\mu \pm \hat\tau)/\sqrt2$; the atmospheric angle is half of a right angle in the literal sense. One $Z_2$ is thus read out twice: its unitary half through moduli, $\theta_{23} = \pi/4$, and its antiunitary half through phases, $\delta = \pm\pi/2$. The two consequences differ in rigidity. The charged-lepton rotation moves $\theta_{23}$ off the bisector linearly in $\theta_e$, but moves the phase off $270^\circ$ only quadratically, Eq.~(\ref{eq:closed23}). Figure~\ref{fig:z2} shows both halves.

\begin{figure}[t]
\centering
\begin{tikzpicture}[scale=1.05]
% ---- left panel: bisector geometry
\begin{scope}[shift={(0,0)}]
\draw[black!50,-{stealth}] (0,0) -- (2.3,0) node[right,font=\scriptsize,black!70]{$\hat\mu$};
\draw[black!50,-{stealth}] (0,0) -- (0,2.3) node[above,font=\scriptsize,black!70]{$\hat\tau$};
\draw[black!70,thick] (0.3,0) -- (0.3,0.3) -- (0,0.3);
\draw[black!45,dashed] (-0.35,-0.35) -- (2.35,2.35);
\node[font=\scriptsize,black!55,rotate=45] at (2.02,2.28) {mirror};
\draw[{stealth}-{stealth},figgreen,thin] (2.05,0.28) to[bend right=28] (0.28,2.05);
\node[font=\scriptsize,figgreen] at (1.72,1.72) {$\mu\leftrightarrow\tau$};
\draw[figblue,line width=1.2pt,-{stealth}] (0,0) -- (1.56,1.56);
\node[font=\scriptsize,figblue,anchor=east] at (1.30,1.48) {$\nu_3$};
\draw[black!70] (0.85,0) arc (0:45:0.85);
\node[font=\scriptsize] at (1.22,0.42) {$\theta_{23}=\tfrac{\pi}{4}$};
\draw[figred,line width=0.8pt,dash pattern=on 4pt off 2pt,-{stealth}] (0,0) -- (37:1.72);
\node[font=\scriptsize,figred,anchor=west] at (37:2.06) {$\theta_e$ (linear)};
\end{scope}
% ---- center panel: fourth-root phase clock
\begin{scope}[shift={(4.9,1.1)}]
\def\r{1.15}
\draw[black!60] (0,0) circle (\r);
\draw[black!30,thin] (-\r-0.2,0) -- (\r+0.2,0);
\draw[black!30,thin] (0,-\r-0.2) -- (0,\r+0.2);
\fill[white,draw=black!70] (\r,0) circle (2.1pt);
\fill[white,draw=black!70] (-\r,0) circle (2.1pt);
\fill[figblue] (0,\r) circle (2.4pt);
\fill[figblue] (0,-\r) circle (2.4pt);
\node[font=\scriptsize,anchor=west] at (\r+0.08,0) {$+1$};
\node[font=\scriptsize,anchor=east] at (-\r-0.08,0) {$-1$};
\node[font=\scriptsize,figblue,anchor=south] at (0,\r+0.08) {$+i$};
\node[font=\scriptsize,figblue,anchor=north] at (0,-\r-0.08) {$-i$};
\node[font=\scriptsize] at (0,\r+0.62) {$e^{4i\varphi}=1$};
\draw[figblue,line width=1.1pt,-{stealth}] (0,0) -- (0,-\r+0.06);
\node[font=\scriptsize,figblue,align=center,anchor=north] at (0.02,-\r-0.42)
  {$\kappa_3=-1$:\ $\delta=\pm\tfrac{\pi}{2}$\\ (shifts only at $O(\theta_e^2)$)};
\end{scope}
% ---- right panel: the alignment director and the half-angle map
\begin{scope}[shift={(9.9,1.0)}]
\draw[black!50,-{stealth}] (-1.55,0) -- (1.7,0) node[above,font=\scriptsize,black!70,yshift=1pt]{$\tfrac{M_{22}-M_{33}}{2}$};
\draw[black!50,-{stealth}] (0,-1.35) -- (0,1.45) node[above,font=\scriptsize,black!70]{$M_{23}$};
\draw[figgreen,line width=1.2pt] (150:1.4) -- (-30:1.4);
\node[font=\scriptsize,figgreen,anchor=north west] at (-30:1.32) {$\alpha_\phi=-\tfrac{\pi}{6}$};
\draw[figblue,line width=1.2pt,{stealth}-{stealth}] (165:1.55) -- (-15:1.55);
\node[font=\scriptsize,figblue,anchor=west] at (1.62,-0.40) {$\theta=-\tfrac{\pi}{12}$};
\draw[black!60] (0.85,0) arc (0:-30:0.85);
\draw[black!60,dotted] (1.1,0) arc (0:-15:1.1);
\node[font=\scriptsize,align=center,black!70] at (0.15,-1.75) {eigenframe at $\alpha_\phi/2$};
\end{scope}
\end{tikzpicture}
\caption{One $Z_2$, read twice. Left: the unitary half of the $\mu$--$\tau$ reflection exchanges the flavor axes; its mirror line bisects their right angle, and the invariant mass eigenstate $\nu_3$ is pinned to the bisector, $\theta_{23} = \pi/4$, half of a right angle. The charged-lepton rotation moves $\nu_3$ off the bisector linearly in $\theta_e$ (dashed). Center: the antiunitary half quantizes the electron-row phases to fourth roots of unity, $e^{4i\varphi} = 1$; columns with $\kappa_i = +1$ take the real pair (open), the third column takes the imaginary pair (filled), giving $\delta = \pm\pi/2$, and the phase moves off $270^\circ$ only at second order in $\theta_e$, Eq.~(\ref{eq:closed23}). Right: the alignment origin of the $\pi/12$ (Sec.~\ref{sec:origin}); in the 2--3 block of the Majorana matrix, the traceless perturbation set by $\langle\phi\rangle$ is a director at angle $\alpha_\phi = -\pi/6$ in the plane of the diagonal difference and the off-diagonal entry, two ticks of the auxiliary group's $e^{i\pi/12}$ phases, and the eigenframe turns at half the director angle, $\theta = \alpha_\phi/2 = -\pi/12$. Every angle of the model is a root of unity read through a half-angle map.}
\label{fig:z2}
\end{figure}
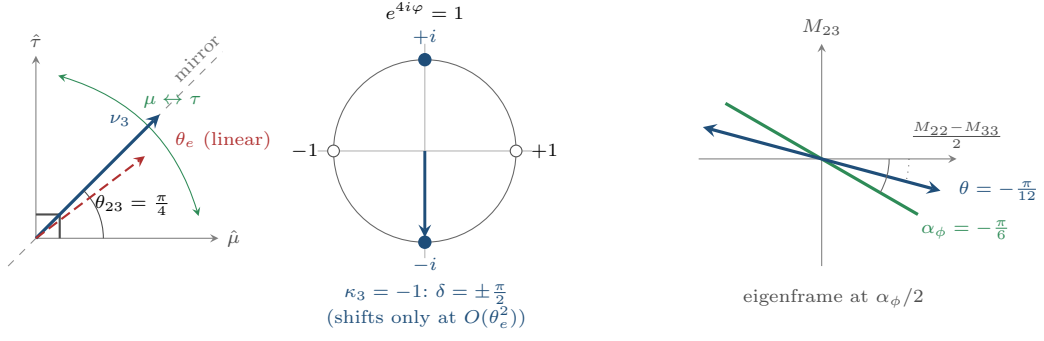

\emph{The dressed matrix.} In the real-rotation limit the dressed matrix, Eq.~(\ref{eq:dressed}) with $\alpha = 0$, takes the factorized form
\begin{equation}
\setlength{\arraycolsep}{9pt}
U(\theta_e) =
\begin{pmatrix}
\sqrt{\dfrac{2}{3}} & \dfrac{c_\pi}{\sqrt3} & i\,\dfrac{s_\pi}{\sqrt3}\\[8pt]
-\dfrac{p}{\sqrt6} & p\,\dfrac{c_\pi}{\sqrt3} + i\,q\,\dfrac{s_\pi}{\sqrt2} & q\,\dfrac{c_\pi}{\sqrt2} + i\,p\,\dfrac{s_\pi}{\sqrt3}\\[8pt]
-\dfrac{q}{\sqrt6} & q\,\dfrac{c_\pi}{\sqrt3} - i\,p\,\dfrac{s_\pi}{\sqrt2} & -p\,\dfrac{c_\pi}{\sqrt2} + i\,q\,\dfrac{s_\pi}{\sqrt3}
\end{pmatrix},
\qquad
\begin{aligned}
p &\equiv \cos\theta_e - \sin\theta_e,\\
q &\equiv \cos\theta_e + \sin\theta_e,
\end{aligned}
\label{eq:Udressed}
\end{equation}
\end{widetext}
with $p^2 + q^2 = 2$ and $q^2 - p^2 = 2\sin 2\theta_e$. The single unknown enters through $(p, q)$ alone: the electron row carries neither, every real part of the $\mu$--$\tau$ block scales with one of the pair and every imaginary part with the other, in the paired combinations $c_\pi/\sqrt3 \leftrightarrow s_\pi/\sqrt2$ and $c_\pi/\sqrt2 \leftrightarrow s_\pi/\sqrt3$, and the exchange $p \leftrightarrow q$ is the $\mu$--$\tau$ exchange, so $\theta_e = 0$ restores the reflection. Row unitarity holds for each row via $p^2 + q^2 = 2$. For general $\alpha$ the $\mu$ and $\tau$ rows are $c_e\,\mathbf{m} - s_e e^{-i\alpha}\,\mathbf{t}$ and $s_e e^{i\alpha}\,\mathbf{m} + c_e\,\mathbf{t}$, with $\mathbf{m}$, $\mathbf{t}$ the second and third rows of Eq.~(\ref{eq:Unu}).

\emph{Observables.} Reading off the observables from Eq.~(\ref{eq:Udressed}):
\begin{align}
|U_{e3}|^2 &= \frac{s_\pi^2}{3} = \frac{2-\sqrt3}{12}, \qquad
\varepsilon_2 = \frac{q^2 - p^2}{2} = \sin 2\theta_e,\nonumber\\
\stt &= \frac{12}{10+\sqrt3}\left[\frac{2+\sqrt3}{8}\,q^2 + \frac{2-\sqrt3}{12}\,p^2\right]\nonumber\\
&= \frac12 + \frac{2+5\sqrt3}{2(10+\sqrt3)}\,\sin 2\theta_e,\nonumber\\
J &= \mathrm{Im}\!\left(U_{\mu3}U^*_{e3}U_{e2}U^*_{\mu2}\right)\nonumber\\
&= -\frac{pq}{12\sqrt6} = -\frac{\cos 2\theta_e}{12\sqrt6}.
\label{eq:appobs}
\end{align}
The linear response of $\stt$ and the quadratic protection of $J$ quoted in Eq.~(\ref{eq:closed23}) follow directly, and $\theta_e = 0$ recovers $J = -1/(12\sqrt6)$ of Ref.~\cite{Krishnan:2019}. The structure of these coefficients traces to the electron row: the Jarlskog invariant factorizes as $J = (|U_{e2}||U_{e3}|)\, c_{12}c_{13}\, s_{23}c_{23}\sin\delta$, and with the exact product $|U_{e2}||U_{e3}| = 1/12$ of Sec.~\ref{sec:erow} and $|U_{e1}| = c_{12}c_{13} = \sqrt{2/3}$ this becomes $J = \sin 2\theta_{23}\sin\delta/(12\sqrt6)$; the electron row fixes the magnitude of leptonic CP violation, and only the factor $\sin 2\theta_{23}\sin\delta = -\cos 2\theta_e$ is left for the $\mu$--$\tau$ sector to supply.

\emph{Numerical evaluation.} Evaluating Eq.~(\ref{eq:Udressed}) at the fitted $\theta_e = -2.23^\circ$ of Eq.~(\ref{eq:fitpq}) gives the complete numerical mixing matrix of the $\pi/12$ model, in the phase convention of Eq.~(\ref{eq:Udressed}),
\begin{equation}
|U| =
\begin{pmatrix}
0.8165 & 0.5577 & 0.1494\\
0.4238 & 0.6050 & 0.6740\\
0.3921 & 0.5683 & 0.7234
\end{pmatrix},
\label{eq:Unum}
\end{equation}
\begin{equation}
\arg U =
\begin{pmatrix}
0 & 0 & +90.00\\
180 & +16.89 & +13.31\\
180 & -19.53 & +168.56
\end{pmatrix}\!{}^\circ,
\label{eq:Uarg}
\end{equation}
unitary to machine precision. The corresponding angles and phases of the $\pi/12$ model are
\begin{align}
\theta_{12} &= 34.33^\circ, & \stw &= 0.31811,\nonumber\\
\theta_{13} &= 8.59^\circ, & \sth &= 0.02233,\nonumber\\
\theta_{23} &= 42.97^\circ, & \stt &= 0.46467,\nonumber\\
\delta &= 271.9^\circ, & (\alpha_{21},\alpha_{31}) &= (0,\ \pi),
\label{eq:finalangles}
\end{align}
with $J = -0.0339$ and $\eps{2} = -0.078$; the Majorana phases are quoted in the PDG convention $P = \mathrm{diag}(1, e^{i\alpha_{21}/2}, e^{i\alpha_{31}/2})$. Individual element phases are convention-dependent; only rephasing-invariant combinations, such as the quartet phase $\arg(U_{e1}U_{\mu2}U^*_{e2}U^*_{\mu1}) = -163.1^\circ$, carry physical meaning. The Majorana phase $\alpha_{31} = \pi$, equivalently the matrix $\mathrm{diag}(1,1,i)$ on the right of Eq.~(\ref{eq:Unum}), encodes the sign pattern $\eta = (+,+,-)$ of Sec.~\ref{sec:tests} and is what enters $m_{\beta\beta}$.

\section{Alignment origin of the real rotation}
\label{app:flavon}

This appendix collects the alignment algebra behind Sec.~\ref{sec:origin}. Throughout,
\begin{equation}
U_\omega = \frac{1}{\sqrt3}
\begin{pmatrix}
1 & 1 & 1\\
1 & \omega & \bar\omega\\
1 & \bar\omega & \omega
\end{pmatrix},
\qquad \omega = e^{2\pi i/3},
\label{eq:Uomega}
\end{equation}
is the trimaximal matrix that left-diagonalizes the charged-lepton mass matrix of Ref.~\cite{Krishnan:2019}, whose columns are set by $\langle\chi\rangle \propto (1,\omega,\bar\omega)$, $\langle\chi^*\rangle \propto (1,\bar\omega,\omega)$, and $\langle(\chi^*\chi)_{\bm{3'}}\rangle \propto (1,1,1)$ coupling to $\tau_R$, $\mu_R$, and $e_R$ respectively.

\emph{Induced rotation.} A flavon $\xi$ contributing to the $\tau_R$ column perturbs the third column of $M_l$ by a vector $\propto \langle\xi\rangle$. In the $U_\omega$ basis the perturbation has components $[U_\omega\langle\xi\rangle]_i$, with $[X]_i$ again the $i$th component of the vector $X$; the entry $i=2$, complex in general, mixes the $\mu$ and $\tau$ left-handed states and induces
\begin{equation}
\theta_e\, e^{i\alpha} \;\propto\; \frac{[U_\omega\langle\xi\rangle]_2}{m_\tau/(v\,v_\chi/\Lambda)} ,
\end{equation}
so the phase $\alpha$ is the phase of $[U_\omega\langle\xi\rangle]_2$, fixed by the alignment of $\xi$ rather than by a coupling. For the candidate directions,
\begin{equation}
\begin{array}{lcl}
\langle\xi\rangle \propto (1,1,1) &:& [U_\omega\langle\xi\rangle]_2 = 0,\\
\langle\xi\rangle \propto (1,\omega,\bar\omega) &:& [U_\omega\langle\xi\rangle]_2 = 0,\\
\langle\xi\rangle \propto (0,1,-1) &:& [U_\omega\langle\xi\rangle]_2 = i \ \ (\alpha = \tfrac{\pi}{2}),\\
\langle\xi\rangle \propto (1,\bar\omega,\omega) &:& [U_\omega\langle\xi\rangle]_2 = \sqrt3 \ \ (\alpha = 0).
\end{array}
\label{eq:aligntable}
\end{equation}
The first two directions produce no rotation; the third produces a purely imaginary entry, $\cos\alpha = 0$, hence $\eps{2} = 0$ and no observable effect on the first-column balance. The conjugate-trimaximal direction is the unique candidate that yields a real rotation, and it does so with the same phase structure as the $\tau$ mass itself, $[U_\omega\langle\chi\rangle]_3 = \sqrt3$.

\emph{Self-protection of the undressed model.} The three bilinears of $\chi$ in the $\bm{3'}$ channel evaluate, with the tensor product $(ab)_{\bm{3'}} = (a_2b_3 + a_3b_2,\, a_3b_1 + a_1b_3,\, a_1b_2 + a_2b_1)$, to
\begin{equation}
\begin{array}{lclcl}
(\chi^*\chi^*)_{\bm{3'}} &\propto& (1,\omega,\bar\omega), && C_3\ \text{charge}\ \omega,\\
(\chi\,\chi)_{\bm{3'}} &\propto& (1,\bar\omega,\omega), && C_3\ \text{charge}\ \bar\omega,\\
(\chi^*\chi)_{\bm{3'}} &\propto& (1,1,1), && C_3\ \text{charge}\ 1.
\end{array}
\label{eq:bilinears}
\end{equation}
Each bilinear carries exactly the $C_3$ charge that routes it into the column whose alignment it reproduces: $(\chi^*\chi^*)$ can couple only to $\tau_R$ and points along $\langle\chi\rangle$; $(\chi\chi)$ only to $\mu_R$, along $\langle\chi^*\rangle$; $(\chi^*\chi)$ only to $e_R$, along $(1,1,1)$. Higher-order operators built from $\chi$ alone therefore renormalize the Yukawa couplings without rotating the left-handed basis, to all orders in these bilinears. The rotation of Sec.~\ref{sec:model} cannot arise as a radiative or higher-dimensional artifact of the existing flavon content; a genuinely new field with the $(1,\bar\omega,\omega)$ alignment is required, which is why we regard $\theta_e \neq 0$ as a falsifiable structural statement rather than an inevitable correction.

\emph{Charges and size.} The operator
\begin{equation}
y_\xi\, \bar L\, \frac{\rho\,\xi}{\Lambda^2}\, \tau_R H
\label{eq:xiop}
\end{equation}
is invariant provided $\xi$ carries the $C_3$ charge $\omega$ of $\chi$ (routing it to $\tau_R$), odd $C_2$ parity to compensate $\rho$, and the appropriate $C_4$ assignment; charge and alignment are independent data, so the assignment $\{\text{charge of }\chi,\ \text{alignment of }\chi^*\}$ is consistent. The induced angle is
\begin{equation}
\theta_e \sim \frac{y_\xi}{y_\tau}\,\frac{v_\rho}{\Lambda}\,\frac{v_\xi}{v_\chi}
\approx 0.7^\circ \times \frac{y_\xi}{y_\tau}\,\frac{v_\xi}{v_\chi},
\end{equation}
using the model's expansion parameter $v_\rho/\Lambda \approx 0.012$. The fitted $\theta_e = -2.23^\circ$ of Eq.~(\ref{eq:fitpq}) requires a combined coefficient near $3.2$, comparable to the $y_\mu \approx 2.4$ the model already carries. \emph{The $\xi$ potential.} The charges of $\xi$ are fixed by Eq.~(\ref{eq:xiop}) to $(\bm{3'},\,1,\,\omega,\,-1)$ under $S_4 \times C_4 \times C_3 \times C_2$. The $C_2$ parity forbids the cubic invariant $(\xi(\xi\xi)_{\bm{3'}})_{\bm 1}$ and the quadratic mixing $(\xi^\dagger\chi)_{\bm 1}$, and the $C_3$ charge forbids all non-Hermitian terms in $\xi$ alone, so the renormalizable potential consists of $m_\xi^2\,\xi^\dagger\xi$, Hermitian quartic self-couplings, the four cross-channel invariants $|(\xi^\dagger\chi)_{\bm r}|^2$ with $\bm r = \bm 1, \bm 2, \bm 3, \bm{3'}$, the pair $[(\xi\xi)_{\bm r}(\chi^\dagger\chi^\dagger)_{\bm r}]_{\bm 1} + \mathrm{h.c.}$, and the cubic $\rho\,(\xi^\dagger\chi)_{\bm 1} + \mathrm{h.c.}$ The antisymmetric channel does the selecting, by the Lagrange identity
\begin{equation}
|(\xi^\dagger\chi)_{\bm 3}|^2 = (\xi^\dagger\xi)(\chi^\dagger\chi) - |(\xi\chi)_{\bm 1}|^2 ,
\label{eq:lagrange}
\end{equation}
which vanishes if and only if $\xi \propto \chi^*$: a positive coefficient on this single invariant costs nothing at the conjugate-trimaximal direction and is strictly positive everywhere else. We have confirmed by global numerical minimization over $\xi \in \mathbb{C}^3$ that $\langle\xi\rangle = v_\xi(1,\bar\omega,\omega)$ is the exact global minimum in an open region of the renormalizable couplings (roughly three quarters of randomly sampled symmetry-breaking points in our scan). Qualitatively, the target is the global minimum whenever the cross couplings $\kappa_1$ and $\kappa_3$ (the coefficients of $|(\xi^\dagger\chi)_{\bm 1}|^2$ and $|(\xi^\dagger\chi)_{\bm 3}|^2$) are not small compared with the self-couplings that favor competing orbits. The one $U(1)_\xi$-breaking invariant, $[(\xi\xi)_{\bm r}(\chi^\dagger\chi^\dagger)_{\bm r}]$, deforms the minimum by an admixture of the $(1,1,1)$ direction, linear in its coupling; the deformation is inert, since $[U_\omega(1,1,1)]_2 = 0$, and numerically $\alpha = 0$ holds to machine precision across the scanned region, so the real-rotation limit is exact, not approximate, at the minimum. That statement takes the potential couplings real, as the generalized CP of the construction dictates; a complex coefficient on $[(\xi\xi)_{\bm r}(\chi^\dagger\chi^\dagger)_{\bm r}] + \mathrm{h.c.}$ would feed the deformation into $\alpha$. The cubic $\rho\,(\xi^\dagger\chi)_{\bm 1}$ vanishes at the vacuum, where $(\xi^\dagger\chi)_{\bm 1} = 0$; its tadpoles admix $\langle\chi\rangle$ and $\langle\xi\rangle$ into one another along directions that are real in the $U_\omega$ basis, renormalizing $m_\tau$ and $\theta_e$ without generating a phase. Back-reaction on the neutrino sector proceeds only through the neutral quartics $(\xi^\dagger\xi)_{\bm r}(\phi^\dagger\phi)_{\bm r}$ and $(\xi^\dagger\xi)(s^* s)$; at the vacuum $(\xi^\dagger\xi)_{\bm{3'}} \propto (1,1,1)$, the same tilt operator that $(\chi^\dagger\chi)_{\bm{3'}} \propto (1,1,1)$ already applies to $\phi$ in the base model, so $\xi$ introduces no new class of alignment perturbation, and taking $\xi$ neutral under the auxiliary group leaves the $Y_{24}$ construction that fixes $\langle\phi\rangle$ untouched. The bookkeeping of the recast model completes the demonstration. In the auxiliary-group version of Ref.~\cite{Krishnan:2019}, the flavour group is extended to $(S_4 \times C_4 \times C_3 \times C_2) \times (Y_{24} \rtimes C_2 \times C_2 \times C_2)$, the triplet $\phi$ is replaced by the effective combination $(\acute\Phi\,\Delta\,\grave\Phi)_{\bm{3'}}/\Lambda^2$ with the elementary flavons carrying the $Y_{24}$ and auxiliary-$C_2$ charges, and $\chi$ is a $Y_{24}$ singlet with trivial auxiliary charges, so the $\xi$--$\chi$ selection sector above carries over verbatim. Assigning $\xi$ a singlet of the auxiliary group (invariant, like $\chi$, under its conjugation $C_2$), the charge arithmetic closes the operator inventory: since the auxiliary-sector fields are all $C_3$-neutral, every operator linear in $\xi$ requires a $\chi^\dagger$, and since $\rho$ and $\xi$ are the only $C_2$-odd scalars, it also requires exactly one $\rho$, reducing every such term to an auxiliary-dressed version of the cubic $\rho\,(\xi^\dagger\chi)_{\bm 1}$, which vanishes at the vacuum; direct mixing of $\xi$ with the effective triplet is forbidden twice over, by $C_3$ ($\omega \neq 1$) and by $C_4$ (the effective triplet carries $-1$, $\xi$ carries $+1$); and every remaining coupling is of the neutral-pair class $(\xi^\dagger\xi)_{\bm r}(X^\dagger X)_{\bm r}$ with $X \in \{\acute\Phi, \grave\Phi, \Delta, \hat s_a, \hat s_b, S\}$, the same class the recast model already contains with $\chi^\dagger\chi$ in place of $\xi^\dagger\xi$. The auxiliary sector therefore adds nothing beyond the charge-allowed set, and the demonstration is complete at the operator level. Only the joint minimization of the full multi-flavon potential across sectors lies beyond our scope, as it does for the base model itself.

\end{document}